\documentclass[12pt]{article}
\usepackage[dvips]{graphicx}

\catcode`\@=11

\def\v#1{\mathchoice{{\mathord{\mbox{\boldmath$\displaystyle #1$}}}}{{\mathord{\mbox{\boldmath$\textstyle #1$}}}}{{\mathord{\mbox{\boldmath$\scriptstyle #1$}}}}{{\mathord{\mbox{\boldmath$\scriptscriptstyle #1$}}}}}

\def\malignl#1{\vcenter{\openup1\jot\ialign{$\displaystyle##{}$\hfil&$\displaystyle{}##{}$&&$\displaystyle##$\hfil\cr#1\crcr}}}
\def\fd{^{\vphantom{\dagger}}}

\def\@makecaption#1#2{\medskip \centering \footnotesize \noindent {\bf #1} \par}

\begin{document}

\title{Coupled electricity and magnetism in solids: multiferroics and beyond}
%\dummyfootnote{An introduction chapter for the planned book on multiferroics,  ed.\ J.~L.~Wang}}
\author{D. I. Khomskii\\[\medskipamount]
\normalsize II. Physikalisches Institut, Universit\"at zu K\"oln,\\
\normalsize Z\"ulpicher Str.\ 77,
\normalsize 50937 K\"oln, Germany}
\date{}

\maketitle

\begin{abstract}
\noindent
The interplay of electricity and magnetism, one of the cornerstones of modern physics,
takes a special form in solids in such phenomena as magnetoelectricity
and the possibility of multiferroic behaviour. In this paper
I give a short survey of the main notions of this field, paying special
attention to microscopic aspects. Some related phenomena, such as
electric activity of magnetic domain walls, etc., are also shortly discussed.
\end{abstract}

\section{Introduction}
\label{sec:1}
The intrinsic coupling of electricity and magnetism is one of the
cornerstones of modern physics. It goes back to the famous Maxwell
equations, or even earlier, to Michael Faraday, and one
can find even earlier reports pointing in that direction. This
coupling plays crucial role in all modern physics, and it is one
of the foundations of modern technology --- e.g.\ in the generation of
electricity in electric power stations, electric transformers, etc.
Recently this field acquired new life in spintronics,
the idea of which is to use not only
charge, but also spin of electrons for electronic applications. Mostly one deals in this field
with the influence of magnetic field and/or magnetic ordering on
transport properties of materials --- for example the well-known
magnetoresistance or the work of magnetic tunnel junctions. But very
interesting such effects can also  exist  in insulators. These are
for example the ({\it linear}) {\it magnetoelectric} (ME) {\it effect}, or the
coexistence and  mutual influence of two types of ordering, magnetic
and ferroelectric (FE) ordering in {\it multiferroics}~(MF). Such phenomena
are very interesting physically, and are very promising for practical
applications, e.g.\ for addressing magnetic memory electrically
without the use of currents, or as very efficient magnetic
sensors. These factors probably caused such a significant interest in
this field. It is now one of the hottest topics in condensed matter
physics, and, besides magnetoelectrics and multiferroics per se, the
study of these has many spin-offs in the related fields of physics,
such as the study of magnetoelectric effects in different magnetic
textures (domain walls, magnetic vortices, skyrmions etc.)

There are already several good reviews of this field
\cite{fiebig,cheong-mostovoy,khomskiiJMMM,ehrenreich,kitaicy,nagaosa,khomskii-trends},
and there
exists a very complete and useful collection of short reviews on multiferroics in
the special issue of Journal of Physics of Condensed Matter~\cite{JPCM}.
There is also a chapter on multiferroics in my recent book~\cite{khomskiiTM}. In
the present text, which is written as an introductory chapter for
a planned book on multiferroics, I will more or less
follow the general outline of my short review on ``Multiferroics for
pedestrians'', published in Physics: Trends~\cite{khomskii-trends} --- of course with significant additions.

Multiferroics are materials with
coexisting magnetic and ferroelectric ordering.
These systems are extremely interesting physically,
and they promise many important practical applications.
However one has to
realize that for many practical applications, such as attempts to
write and read magnetic memory in hard discs electrically,
% i.e.\
using
electric fields rather than currents (e.g.\ with gate voltage devices), one needs
not so much multiferroics but rather materials with good
magnetoelectric properties: one must be able to {\it modify} the magnetic
state by a changing electric field.
%From this point of view one may not even need real multiferroics, but rather good magnetoelectrics.
But the idea is that it is precisely
multiferroic materials in which the change of magnetic state by electric field, or vice
versa, may be especially strong. From this point of view,
various textures in magnetic materials
which can have magnetoelectric response --- such as certain domain walls of skyrmions --- also attract
now considerable attention.
These topics will be also mentioned below.
%, and extensively discussed in several chapters of this book.

\section{Some historical notes}
\label{sec:2}
When one describes the field of magnetoelectrics and multiferroics,
the first reference one usually gives is that to Pierre Curie~\cite{curie},
who shortly noticed the possibility of having
both magnetic and electric orderings in one material. But the real story
began with a short remark in one of the famous books on
theoretical physics by Landau and Lifshitz~\cite{landaulifshitz}, who wrote in 1959:
\begin{quotation}
\noindent``Let us point out two more phenomena, which, in principle,
could exist. One is piezomagnetism, which consists of linear
coupling between a magnetic field in a solid and a deformation
(analogous to piezoelectricity). The other is a linear coupling
between magnetic and electric fields in a media, which would
cause, for example, a magnetization proportional to an electric
field. Both these phenomena could exist for certain classes
of magnetocrystalline symmetry. We will not however discuss
these phenomena in more detail because it seems that till
present, presumably, they have not been observed in any substance.''
\end{quotation}
Indeed, at the moment of publication of that volume there were no
known real examples of magnetoelectric or multiferroic systems. But already less than a year after
its publication the seminal paper by Dzyaloshinskii appeared~\cite{dzyaloshinskii},
who on symmetry grounds predicted that the well-known antiferromagnet
Cr$_2$O$_3$ should exhibit the linear ME effect. And next year this effect was
indeed observed in Cr$_2$O$_3$ by Astrov~\cite{astrov}. After that a rapid
development of this field followed, initially in the study of magnetoelectrics, see e.g.~\cite{freeman}.
But very soon the ideas of not only the ME effect, but of real
multiferroics were put forth. Soon the first
multiferroic --- a material in which
(antiferro)magnetic and ferroelectric ordering are present simultaneously --- was discovered
by Ascher, Schmid et~al.\
\cite{schmid-boracite} -- the Ni--I boracite. (It was in fact Hans Schmid who later
coined the very term ``multiferroics'' in connection with such
materials~\cite{schmid-term}). An active program to synthesize such
materials artificially was initiated, predominantly by two groups in the former
Soviet Union: in the group of Smolenskii in Leningrad (present-day
St.~Petersburg) and in the group of Venevtsev in Moscow.

However, after considerable activity in the 1960s and 1970s,
the interest in this field faded somewhat.
A new surge of activity appeared at around
2000, and there were three factors which stimulated it:

The first was the realization of an interesting and challenging
problem in the physics of magnetic and ferroelectric materials,
mostly on the example of perovskites. There are quite a lot of
magnetic perovskites, including famous colossal magnetoresistance
manganites, or the ``two-dimensional perovskite'' La$_2$CuO$_4$ --- the parent
material of High-Temperature superconducting cuprates.  Extensive
discussion of these materials, with many tables, is contained in the
collection compiled by Goodenough and Longo in the Landolt-B\"ornstein
Encyclopaedia of Physics~\cite{goodenough-longo}. Another, even more extensive collection
of tables of ferroelectric perovskites, starting with the equally
famous material BaTiO$_3$, was published
%in the same edition
by a group of
Japanese scientists~\cite{landolt}. And, surprisingly enough, a comparison of
these extensive collections of tables, 100--300 pages each,
demonstrates that there is practically no overlap between them: a
perovskite is either magnetic or ferroelectric, but practically never
both simultaneously (of course it may be neither, as is
the case with the prototype mineral perovskite CaTiO$_3$, which gave the name to this
whole family). What is the reason for this mutual exclusion, and can one go around
it? This problem was known already in 1970--1980s, but was actually
formulated only in 1999 during a workshop in Santa Barbara, and
publicised after 2000~\cite{APS,hill}, and it attracted the attention of
scientific community. This problem will be discussed below, in Sec.~\ref{sec:6}.

But of course the most important were two experimental breakthroughs.
One was the fabrication and study of very good films of, it
seems, the best multiferroic material known at present, BiFeO$_3$, by
Ramesh and his group~\cite{wang}. This gave a possibility of studying the MF
effects, and immediately opened perspectives of very appealing
practical applications. BiFeO$_3$ remains until today the favourite
material for many investigations, both in basic research and in
applied fields.

The second achievement was the discovery by two groups, of Kimura and
Tokura, and of Sang-Wook Cheong, of a novel class of multiferroics
\cite{kimura,cheong}.
In multiferroics that were known previously the ferroelectric and magnetic
orderings occurred independently and were driven by different mechanisms;
typically, although not always, FE ordering starts at higher
temperature. In the novel class of multiferroics discovered in \cite{kimura,cheong}, FE is
driven by a particular type of magnetic ordering, and occurs only in
the magnetically-ordered phase. One can call the first group ``type-I
multiferroics'', and the second group ``type-II multiferroics''~\cite{khomskii-trends}. I will
discuss this classification and the microscopic mechanisms in action in
each of these classes in Sec.~\ref{sec:5}. Here I only want to stress
that these two experimental breakthroughs gave new life to the whole
field of MF and led to an enormous increase of activity in this
field. % --- the present book being a good example of that.

\section{Magnetoelectric effect; symmetry considerations}
\label{sec:3}
The specific feature of magnetoelectric materials is the possibility
of generating electric polarization by magnetic field, and vice versa,
magnetization by electric field. This can be described by the relations
\begin{eqnarray}
P_i &=& \alpha_{ij} H_j + \beta_{ijk} H_jH_k + \cdots \;,\label{eq:1}\\
M_j &=& \alpha_{ji} E_i + \beta_{jik} E_iE_k + \cdots \;,\label{eq:2}
\end{eqnarray}
where we use the standard convention of summation over the repeated
indices.  Terms quadratic in $E$, $H$ are typically more common and less
interesting; the most interesting effect
%to which one usually refers as the ME effect
is the presence of the first, linear terms above.  This is referred to
as the linear ME effect, or simply the ME effect.

One can also describe the linear ME effect by including in the expression
for the free energy the term
\begin{equation}
F_{\rm ME} = -\alpha_{ij} E_iH_j\;.
\label{eq:3}
\end{equation}
As the polarization and the magnetization are given by
$P = - \partial F/\partial E$ and $M=-\partial F/\partial H$,
one immediately obtains from this expression for the free energy the first
terms in (\ref{eq:1}) and~(\ref{eq:2}).

As we see, in general in a crystal the magnetoelectric coefficient~$\alpha$
is a tensor. It can have symmetric and antisymmetric
components. The symmetric part of this tensor can always be
transformed to a diagonal form
\begin{equation}
\alpha_{ij} = \alpha_i \delta_{ij}
\end{equation}
(where $\delta_{ij}$ is the Kronecker symbol). In this case the polarization for
the magnetic field along the main axes would be parallel to the magnetic field.
But there can also be an antisymmetric part of the magnetoelectric
tensor. It is known that such an antisymmetric tensor, with independent components $\alpha_{12}$,
$\alpha_{13}$ and $\alpha_{23}$, is equivalent to an axial vector, or pseudovector
\begin{equation}
T_i = \varepsilon_{ijk} \alpha_{jk}
\label{eq:5}
\end{equation}
where $\varepsilon_{ijk}$ is the totally antisymmetric Levi-Civita symbol. This
pseudovector~$\v T$ is called the toroidal moment.  If the system has a
nonzero toroidal moment, then, from (\ref{eq:1}), (\ref{eq:2}), (\ref{eq:5}) one can see that
for example the polarization in an external magnetic field would be
\begin{equation}
\v P \sim \v T \times \v H\;,
\end{equation}
and magnetization would be
\begin{equation}
\v M \sim \v T \times \v E\;,
\end{equation}
i.e.\ they would be perpendicular to the external fields.

\begin{figure}[ht]
  \centering
  \includegraphics[scale=1]{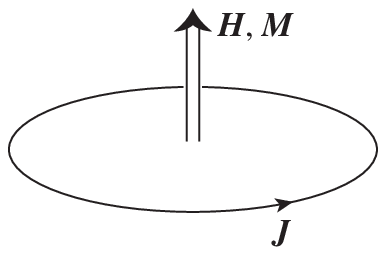}
  \caption{}
  \label{fig:1}
\end{figure}

A very important role in the ME effect is played by symmetry considerations.
First of all, these refer to the symmetry with respect to spatial
inversion,~$\cal J$, and time reversal,~$\cal T$. Electric field, polarization
and electric dipole moments are usual vectors, changing sign under
spatial inversion, but remaining the same under time reversal:
\begin{equation}
\malignl{
  {\cal J}\v P = -\v P\;,      &\qquad&  {\cal J}\v E = -\v E\;,  \cr
  {\cal T}\v P =  \v P\;,      &\qquad&  {\cal T}\v E =  \v E\;.  \cr
}
\label{eq:8}
\end{equation}
On the other hand, magnetization, and the magnetic field itself, are
axial vectors, or pseudovectors, odd with respect to time reversal, but even with respect to
spatial inversion,
\begin{equation}
\malignl{
  {\cal J}\v M =  \v M\;,      &\qquad&  {\cal J}\v H =  \v H\;,  \cr
  {\cal T}\v M = -\v M\;,      &\qquad&  {\cal T}\v H = -\v H\;.  \cr
}
\label{eq:9}
\end{equation}
One can easily understand these rules when one recalls that magnetic
field and magnetic moments are created by currents $\v J = e\v v = e\,d\v r/dt$;
for example, for circular currents shown in Fig.~\ref{fig:1} we have $\v M \sim \v r \times \v J$.
One sees from this expression that $\v M$ is even for spatial
inversion ($\v r\to -\v r$), but odd for time reversal ($t\to-t$).

\begin{figure}[ht]
  \centering
  \includegraphics[scale=1]{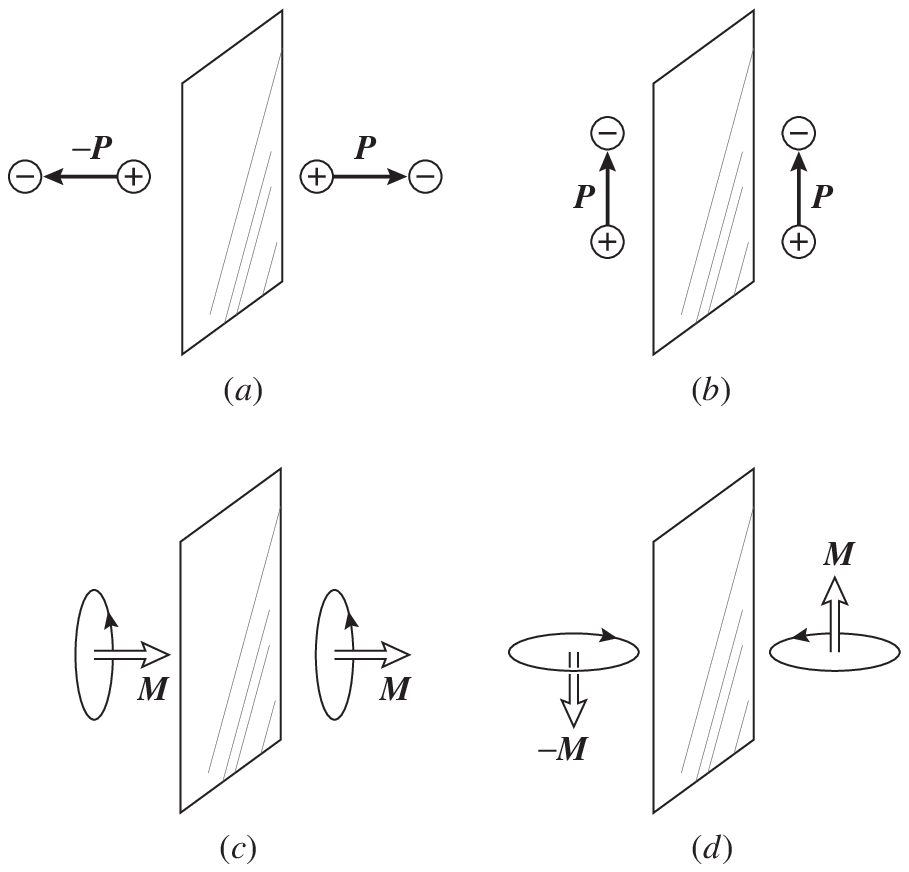}
  \caption{}
  \label{fig:2}
\end{figure}

Using this picture, one can also obtain the transformation rules of
e.g.\ $\v M$ or $\v P$ under mirror reflections, illustrated in Fig.~\ref{fig:2}
(see also~\cite{khomskiiTM}): under mirror reflections the components of $\v P$ and $\v M$ parallel and
perpendicular to the mirror plane change as
\begin{equation}
\malignl{
  P_{\perp} \longrightarrow -P_{\perp}\;,      &\qquad&  P_{\parallel} \longrightarrow  P_{\parallel}\;,  \cr
  M_{\perp} \longrightarrow  M_{\perp}\;,      &\qquad&  M_{\parallel} \longrightarrow -M_{\parallel}\;.  \cr
}
\end{equation}

From the rules (\ref{eq:8}), (\ref{eq:9}) one sees that the linear ME effect (liner
terms in (\ref{eq:1}), (\ref{eq:2}), or (\ref{eq:3})) can exist in a system only if
both inversion and time reversal are simultaneously broken: the
energy (\ref{eq:3}) should be a scalar, and thus the ME coefficient $\alpha$ should
be both $\cal J$-  and $\cal T$-odd. For that, first of all, the system should
have some magnetic ordering which breaks time reversal. In most
cases this is the standard magnetic ordering, for which the average spin at a
site $\langle S_i\rangle \neq 0$, but one cannot exclude more complicated states
such that not the magnetic dipole $\langle S\rangle$ is nonzero, but rather
there exists some non-zero
 higher-order spin correlation function, containing an odd number of
spins --- e.g.\ the magnetic octupole $\sim\langle S_1S_2S_3\rangle$. And
the spatial inversion should also be broken in order to have the linear ME effect; often
this symmetry is broken just by a particular type of magnetic ordering.

In general, the free energy may also contain terms of higher
order, not only those presented in eq.~(\ref{eq:3}). For example,
we may have terms of the type
$\beta_{ijk}E_iH_jH_k$, or similar terms written as a function of order
parameters~$P$ and~$M$, e.g.\ $\sim \beta_{ijk}P_iM_jM_k$, or terms~$\sim P^2M^2$. The
conditions for their appearance are often not so stringent as those
for linear ME coupling. We will not, however, consider such terms
below, and will concentrate on the linear ME effect.

There is one general relation between the ME response function $\alpha_{ij}$
and the usual dielectric and magnetic response, characterized by the
dielectric constant~$\epsilon_{ij}$ (or the corresponding electric
susceptibility, or polarizability), and the magnetic response
characterized by magnetic permeability~$\mu$ or magnetic susceptibility
$\chi$, with $\mu = 1+4\pi\chi$. This constrain has the form~\cite{strickman}
\begin{equation}
\alpha^2 < \chi_{\rm e} \chi_{\rm m}
\end{equation}
where $\chi_{\rm e}$ and $\chi_{\rm m}$ are the electric and magnetic susceptibilities.
We see that one can hope to obtain strong ME coupling for example close
to a ferroelectric or magnetic transition, in which (for II~order transitions)
$\chi_{\rm e}$ or $\chi_{\rm m}$ diverge, $\chi_{\rm e}$ or $\chi_{\rm m} \to \infty$.

One more point is worth addressing here. We now know very well that
the electric and magnetic responses are in general frequency- and
momentum-dependent, $\epsilon(\v q, \omega)$, $\chi(\v q, \omega)$.
This dependence has very definite physical meaning. Thus
for example the dielectric function contains terms such as
\begin{equation}
\epsilon(\v q, \omega) \sim \sum \frac{c_i}{\omega^2 - \omega^2_i(\v q)}
\label{eq:12}
\end{equation}
i.e.\ it has poles at the positions of dipole-active collective
excitations~$\omega_i(\v q)$, for example optical phonons.
These modes give definite signatures e.g.\ in the optical properties
of solids.
Similarly, the structure of $\chi(\v q, \omega)$, which can be measured for
example by magnetic neutron scattering, tells us about magnetic
excitations in the system, such as spin waves with their spectrum
$\omega(\v q)$; and the existence of (strong) maximum of
$\chi(\v q,0)$ at a certain $\v q$-value~$\v q_0$  may be a signature of eventual
magnetic instability of the system, such as the formation of spin density
wave with momentum~$\v q_0$, etc.

One should think that, similarly, the ME response function
$\alpha$ should also have both frequency- and momentum-dependence, $\alpha(\v q, \omega)$.
This question was not, to the best of my knowledge, yet
studied in a general form for ME materials. Apparently the
electromagnons~\cite{electromagnons}
%, to be discussed in Ch.~\#\# of this book,
are related to this question --- they should be the
poles of both $\epsilon$ and $\alpha$, similar to eq.~(\ref{eq:12}). But what could be, for
example, the $q$-dependence of~$\alpha$, what would be its significance,
and how can one measure it, are still open questions. One could
think that there should also be some general relations for $\alpha(\v q,\omega)$
similar to the Kramers-Kronig relations or to the optical sum rule for
$\epsilon(\v q, \omega)$, however I am not aware of any such general treatment
yet (possibly one could find some related results in the
literature, but they are not formulated in this language).

\section{Multiferroics}
\label{sec:4}
\subsection{General considerations}
By multiferroics in a narrow sense we refer to materials having
simultaneously both magnetic and ferroelectric ordering, i.e.\ having
two order parameters $\v M(\v r)$ and $\v P(\v r)$\footnote{Sometimes
one includes among multiferroics also systems with a third type of ordering --- a
ferroelastic one.
%But in the present book, at least in this chapter,
Here, however, we will not consider it.}.
Magnetic ordering could be of different types: ferromagnetic, ferri-
or antiferromagnetic, or it could be of some more complicated type. But for
electric ordering one has in mind a real FE ordering, $\langle P\rangle \neq 0$.
Sometimes in the field of ferroelectricity one also speaks about
antiferroelectrics (AFE), but one has to realize that this notions
has no strict physical meaning. Magnetic transitions, of any kind,
always correspond to symmetry-breaking: going from paramagnetic to
magnetically-ordered state we break at least the time reversal symmetry
(and maybe some spatial symmetries as well). Similarly, FE
transition corresponds to a change of symmetry in the system from
centrosymmetric to noncentrosymmetric one. However, the nominally
antiferroelectric transitions do not necessarily break any symmetry:
one can always formally consider any system as having electric
dipoles, e.g.\ inside a unit cell, pointing in opposite directions. In
this sense any structural transition in a solid is accompanied by
some charge redistribution and could be formally called an AFE\null.
Still, sometimes it can make sense to speak about an AFE transition, if
this structural transition is accompanied by relatively strong anomalies in
the dielectric constant~$\epsilon$; but
one has to realize that this notion has no rigorous meaning. In any
case, in the field of multiferroics we always have in mind the appearance of a
real FE polarization, which is nonzero when averaged over the whole
sample --- although sometimes we also speak about {\it local polarization}.

\begin{figure}[ht]
  \centering
  \includegraphics[scale=1]{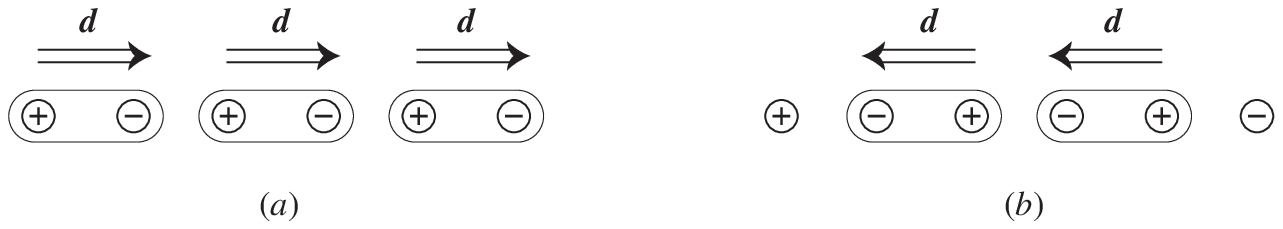}
  \caption{}
  \label{fig:3}
\end{figure}

The different character of electric and magnetic ordering is
reflected in one important aspect. The magnetic order parameter, e.g.\
the magnetization of a ferromagnet~$\langle M\rangle$ or sublattice magnetization
$L = \langle M_1 - M_2\rangle$ of an antiferromagnet are well-defined quantities, having
absolute meaning. This however is not the case with electric polarization: it
may depend for example on the choice of the unit cell, cf.\ Fig.~\ref{fig:3}. It
looks that with the choice of the unit cell as shown in Fig.~\ref{fig:3}($a$) the
polarization points from left to right; however in the same system but
with a different choice of the unit cell, Fig.~\ref{fig:3}($b$), it points from right
to left. And indeed, the accurate treatment shows, see for example
a very pedagogical explanation in~\cite{spaldin}, that the absolute value of
polarization is not a uniquely defined quantity, but {\it the change of
polarization} with changing external conditions, e.g.\ temperature or
electric field, is. This is also reflected in that fact that one has
to use special theoretical methods (the so-called Berry phase methods) for
{\it ab~initio} calculation of polarization.

The very term ``multiferroic'' was proposed by H.~Schmid~\cite{schmid-term}.
In his review article~\cite{schmid-review} Schmid also presented a classification of
different symmetry classes which allow for simultaneous presence of
both FE and magnetic ordering.

\begin{figure}[ht]
  \centering
  \includegraphics[scale=1]{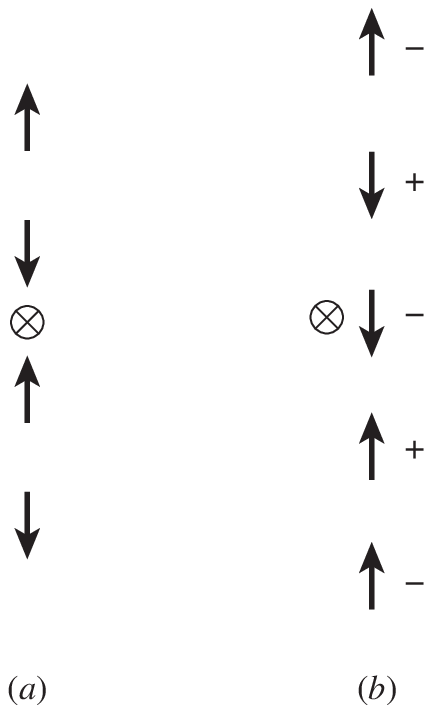}
  \caption{}
  \label{fig:3'}
\end{figure}

In speaking about multiferroics, symmetry considerations play a crucial role.
Both time reversal~$\cal T$ and spatial inversion symmetry~$\cal J$ should
be broken. And one also needs one unique vector~$\v P$ which has to go to
$-\v P$ under inversion. One sees that these symmetry requirements are the
same as those needed to get the linear ME effect. An important question is in
which cases would one get ME, with polarization existing only in an
external (magnetic) field, and when will we have real multiferroics, with
spontaneous polarization existing without any external field. For ME
both $\cal T$ and $\cal J$ should be broken, but the product $\cal TJ$ is
conserved: by consecutive application of time reversal and spatial inversion we
would return to the initial state. For multiferroics, however, not only $\cal T$ and
$\cal J$ but also the product $\cal TJ$ should be broken. Thus by
looking at these symmetries one can understand whether a particular
material with a given magnetic structure would be a real multiferroic or only a magnetoelectric.
The examples mentioned above demonstrate this. The classical ME
material Cr$_2$O$_3$ has a crystal and magnetic structure with the main
element shown in Fig.~\ref{fig:3'}($a$) (the inversion centre is marked by the
encircled cross~$\otimes$). Of course, as in all magnetic states,
time-reversal is broken right away. According to the rules formulated
above, spatial inversion is also broken (it transforms spin $\uparrow$
to~$\downarrow$). But simultaneous inversion and time
reversal (inversion of spin directions) returns the state to the
original one, i.e.\ ${\cal T}|{\rm in}\rangle = -|{\rm in}\rangle$,
${\cal J}|{\rm in}\rangle = -|{\rm in}\rangle$, but
${\cal TJ}|{\rm in}\rangle = |{\rm in}\rangle$. Therefore this system is ME but not MF\null. On the
other hand, for example in the structure shown in Fig.~\ref{fig:3'}($b$), with
alternation of ions with different charges, e.g.\ $+$ and~$-$, and with
the magnetic structure ${\uparrow}\,{\uparrow}\,{\downarrow}\,{\downarrow}$ (this is
a schematic representation of a real situation in Ca$_3$CoMnO$_6$~\cite{choi},
see also~\cite{vdBrink} and Fig.~\ref{fig:10} below),
not only $\cal T$ and $\cal J$, but also $\cal TJ$ are broken,
time reversal following inversion leading to a state different from the
initial one. And indeed this gives real multiferroics~\cite{choi}.

One still has to be slightly careful with his classification: to have a
real FE in the usual sense one has to have the ability
to reverse polarization, $\v P \to -\v P$,
by applying a proper electric field~$E$ (the poling
procedure, which gives the famous polarization loop, $P(E)$ hysteresis).
In some systems, however, the electric field required for that is too
strong to be realized in practice, so that the polarization is always
pointing in one direction, and we cannot switch it. In this case one
speaks not about  ferroelectrics, but rather about pyroelectrics~\cite{landaulifshitz}.
There are such examples also among ``multiferroic'' compounds.
Thus PbVO$_3$ has a tetragonal structure of the same type as the famous
ferroelectric PbTiO$_3$, but the distortion in it leading to the
formation of dipole moments is so strong that the polarization
cannot be reversed~\cite{PbVO3-1,PbVO3-2} (maybe it would possible to realize
this in corresponding films?). Thus PbVO$_3$ should rather be called
a magnetic pyroelectric, not a multiferroic. Still, symmetry
considerations are extremely useful, and actually crucial in the whole
big field of multiferroics. One can find corresponding treatment for example in
the review~\cite{schmid-review} or in~\cite{aharony}. We will not dwell on
this topic any more, and will rather discuss more microscopic aspects of the
physics of multiferroics.

\section{Different types of multiferroics}
\label{sec:5}
Speaking of microscopic mechanisms, one can first of all say that,
despite the huge variety of different types of magnetic ordering,
most of all ``strong'' magnets are conceptually the same, see e.g.~\cite{khomskiiTM}:
due to strong
electron--electron interaction or strong electron correlations the state
is formed with localized electrons (which for
integer number of electrons per cite are the Mott, or
Mott--Hubbard insulators), i.e.\ the state with localized magnetic
moments --- localized spins. The exchange interaction between these
localized moments leads to a certain magnetic ordering at low
temperatures. Depending on specific details, such as electron configuration
of respective ions, orbital occupation, detailed type of the lattice,
etc., we can have quite diverse types of magnetic ordering, but the
general picture --- the presence of localized electrons or localized
spins with particular exchange interaction --- remains the same.

The situation with ferroelectrics is more diverse and much more complicated.
There exist many different microscopic mechanisms leading to ferroelectric
behaviour. And all the diversity of ferroelectrics and of eventual multiferroics is mainly
connected just with this diversity of mechanisms of ferroelectricity.
Thus, we can have systems in which there exist structural
units, e.g.\ molecules similar to HCl, each of which has nonzero dipole
moment just ``by construction''. And some ordered arrangement of such
units could in principle make a material ferroelectric. It is known for example
that among many forms of water ice there is one which is
ferroelectric.

Another mechanism is met in hydrogen-bonded ferroelectrics. To these belong some
inorganic compounds, e.g.\ KH$_2$PO$_4$ (KDP), but mostly organic systems,
for example the first ferroelectric discovered in nature --- the Rochelle or
Seignette's salt NaKC$_4$H$_4$O$_6\cdot4$H$_2$O
(this compound even gave the name to this very phenomenon in
several languages, where ferroelectricity is called
seignetoelectricity) --- although the exact nature of FE in this
material is still a matter of debate.  In both cases above there
may exist some magnetic ions in ``other parts'' of the system, so
that in effect such systems may become multiferroic.

For us, however, other types of multiferroics are of more importance. These are
FE or MF with FE driven by the covalency of transition metal (TM)
ions with surrounding cations (ligands), for example with oxygen;
``geometric'' ferroelectrics; and ferroelectrics with lone pairs. We will now proceed to a short description of
these three classes of materials. However, before that, we
will briefly discuss two general notions.

One can divide all multiferroics into two big groups, which we can call type-I
and type-II multiferroics~\cite{khomskii-trends}; we have already shortly mentioned this
classification in Sec.~2. The multiferroics we have mostly discussed until now, with
the mechanisms of ferroelectric ordering listed above, have independent
mechanisms of FE and magnetic ordering, occurring at different
temperatures (usually with FE transition above the magnetic one, but not
necessarily so). These are type-I multiferroics. Ferroelectricity often occurs in
them at rather high temperatures --- thus in BiFeO$_2$  $T_{\rm FE} \sim 1100\,\rm K$, and in
YMnO$_3$ $T_{\rm FE} \sim 1000\,\rm K$\null. Magnetic ordering, occurring independently, can
also be rather high: in BiFeO$_3$ $T_{\rm N} = 640\,\rm K$, so that it is a good multiferroic
already at room temperature. In general such systems can have quite
large spontaneous polarization, which in BiFeO$_3$ reaches 80--$100\,\rm \mu C/cm^2$
--- larger that in BaTiO$_3$ ($\sim 60\,\rm \mu C/cm^2$). Of course
there also exists a certain coupling between magnetism and ferroelectricity in these
materials, but unfortunately it is usually not strong enough,
although it was demonstrated, for example, that one can modify
the magnetic structure of BiFeO$_3$ by electric field~\cite{viret}.

At the beginning if this century the other, novel class of multiferroics was
discovered~\cite{kimura,cheong} --- the systems which we can call type-II
multiferroics. These are the systems in which ferroelectricity exists and is
generated {\it only} in certain magnetically-ordered states. These
materials attract now the main attention from the point of view of
fundamental science.  They are, however, at least as yet,
less promising for practical applications than type-I systems
such as BiFeO$_3$, or composite multiferroics consisting e.g.\ of layers of
nonmagnetic FE such as (PbZr)TiO$_3$ and adjacent layers of
good magnets such as permalloy, with the coupling between these layers
occurring via common strain (using magnetostriction of magnetic
layers and piezoelectric response of FE ones). But from the physical
point of view these type-II, or magnetically-driven MF, present
special interest\footnote{\label{fn:pfootn}Sometimes one presents a different
classification of MF, paying main attention to the fact whether
polarization is a primary order parameter, or whether it is caused by the
coupling to another one, for example magnetic ordering, e.g.\ due to
coupling $\sim PM^2$ (but possibly also by coupling to some
non-ferroelectric structural distortion).  One calls the first
class of these systems {\it proper} FE, and the second one {\it improper}
FE~\cite{levanyuk}. In the usual Landau theory of II~order phase transitions,
see e.g.~\cite{LLstat,khomskii-aspects}, the primary order parameter~$\eta$ close to the
critical temperature behaves as $\eta \sim \sqrt{T_c-T}$, but for example
for the coupling of the type $P\eta^2$, e.g.\ $PM^2$, polarization is linear in
temperature, $P \sim T_c-T$.}.

\section{Type-I multiferroics}
\label{sec:6}
The varieties of type-I multiferroics differ first of all by the mechanisms
leading to ferroelectricity.  Two such types we have already shortly discussed above.

\medskip

{\it Systems having structural units with permanent dipoles.} To such
systems belong materials containing for example polar groups such as~BO$_3$.
If there are also magnetic ions in such compounds, these could
be multiferroic. Examples thereof are boracites, e.g.\ the Ni--I boracite
Ni$_3$B$_7$O$_{13}$I~\cite{schmid-boracite},
or iron borate $R$Fe$_3$(BO$_3$)$_4$~\cite{zwezdin}.

\medskip

{\it Hydrogen-bonded ferroelectrics.} As mentioned above, to these systems
belong the first known ferroelectric --- the Rochelle (or Seignette's) salt, first
prepared in~1675. There are many materials in this class, but
there are no good multiferroics yet among those.

\medskip

\begin{figure}[ht]
  \centering
  \includegraphics[scale=1]{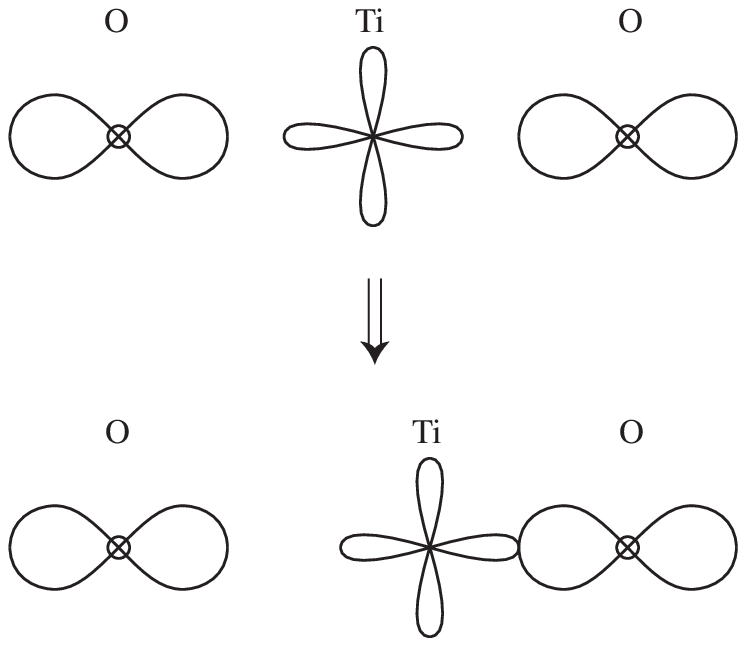}
  \caption{}
  \label{fig:6}
\end{figure}

\looseness=-1
{\it Transition metal perovskites.} Probably the most important class of multiferroics
are the systems such as perovskites in which ferroelectricity is due to
``FE-active'' transition metal (TM) ions. Such are for example the
famous BaTiO$_3$. In classical physics one usually describes FE as a
consequence of the so called polarization catastrophe, at which the
high polarizability of some constituent ions leads to an instability
of the nonpolar state and to creation of ferroelectricity. However there is another,
more microscopic explanation of the appearance of ferroelectricity in systems such as
BaTiO$_3$: the establishment of a strong covalent bond of the transition metal ion, here Ti,
with one (or several) of the surrounding oxygens at the expense of
weakening the bond with the other ones, see Fig.~\ref{fig:6}. One can show, see e.g.~\cite{khomskiiJMMM},
that the energy gain by the corresponding shift
$\delta u$ of Ti ions from the centres of O$_6$ octahedra is $\sim -g(\delta u)^2$.
As the elastic energy loss is also quadratic in~$\delta u$,
$+\frac12B(\delta u)^2$, such phenomenon may occur only if the gain of
covalency energy exceeds the elastic energy loss, for which one needs
strong electron--lattice interaction (large coupling constant~$g$ above)
and a not too stiff lattice (smaller bulk modulus~$B$). These
conditions are met not in all materials even with similar structures.
That is why for example BaTiO$_3$ is ferroelectric; SrTiO$_3$ is not but is ``almost
there'' (it has a very high dielectric constant, and small perturbations
such as uniaxial stress and even isotope substitution $^{16}$O $\to$ $^{18}$O
make it ferroelectric); and CaTiO$_3$ is much further from the ferroelectric
state.

However BaTiO$_3$, a classical ferroelectric, is not magnetic, i.e.\ not a multiferroic. The
analysis of experimental data, e.g.\ the comparison of extensive
tables with the large amount of data on ferroelectric~\cite{goodenough-longo} and magnetic~\cite{landolt}
perovskites shows that usually these two types of ordering in
perovskite family are mutually exclusive. Ferroelectricity in ``classical''
systems is observed practically exclusively in perovskites $AB$O$_3$ with TM
$B$-ions with empty $d$-shells, i.e.\ with the occupation~$d^0$ (BaTi$^{4+}$O$_3$;
LiNb$^{5+}$O$_3$ etc.). On the other hand, magnetism requires partial
occupation of $d$-levels. The realization of this dichotomy caused a
long debate why this is so, and why none (or so few) magnetic perovskites
with $d^n$ ($n \neq 0$) shells are ferroelectric~\cite{APS,hill,khomskiiJMMM}.
There are several physical factors proposed to explain
this property.

One is simply that the ions with empty $d$-shells are usually smaller
than those with $d^n$ ($n\neq0$) configurations, so that such small ions can
easily shift from the centre of a large O$_6$ octahedral cavity.  This
factor may play a certain role, see below, but it does not explain why
for example BaTiO$_3$ is ferroelectric while CaTiO$_3$ is not. Even worse:
CaMnO$_3$ is also not ferroelectric, although it contains ions Mn$^{3+}$ which are
smaller than Ti$^{4+}$ in BaTiO$_3$~\cite{Shannon}.

\begin{figure}[ht]
  \centering
  \includegraphics[scale=1]{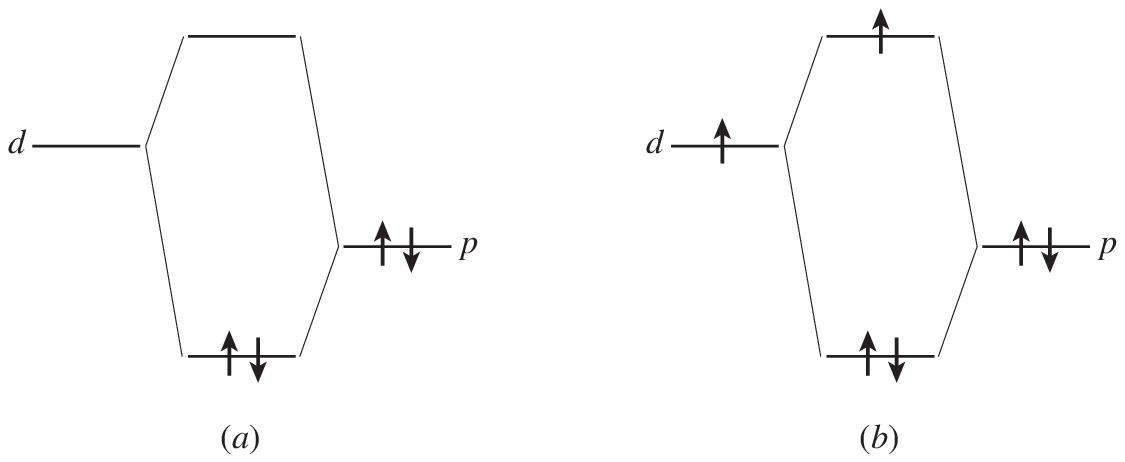}
  \caption{}
  \label{fig:7}
\end{figure}

The other factor could be that, whereas the formation of a strong
covalent bond with, say, one oxygen, see Fig.~\ref{fig:6}, leads to a decrease
of the electron energy (only the bonding orbital is occupied for $d^0$ TM
ions, Fig.~\ref{fig:7}($a$)), in the presence of real $d$-electron(s) on TM the
antibonding orbital should also be filled, Fig.~\ref{fig:7}($b$). Because of that we lose, in this
simple example, half of the energy which we gained by shifting the $d^0$
TM ion.  This effect, if it does not forbid FE in this case, at least makes
it much less probable.

One can think of yet another factor which could suppress the formation of TM--O
covalent bonds leading to ferroelectricity, or at least
make it less favorable~\cite{khomskiiJMMM}. The covalent bond is a typical singlet chemical bond,
e.g.\ with the wave function of the type $\frac{1}{\sqrt2}(d{\uparrow}\,p{\downarrow} -
d{\downarrow}\,p{\uparrow})$, as in the H$_2$ molecule in the Heitler--London approximation. However
in the presence of another localised $d$-electron (or several of those),
as in Fig.~\ref{fig:7}($b$), the Hund's rule exchange would destabilize the singlet
covalent bond --- similarly to the magnetic pair-breaking of singlet
Cooper pairs in $s$-wave superconductors with magnetic impurities~\cite{abrikosov}.

All these factors could be at work and make the coexistence of
magnetism and ferroelectricity in these systems very unlikely. But one has to
realize that this mutual exclusion is not a ``theorem''; it is a game
of numbers, and the physical factors described above, though
making this coexistence highly unlikely, do not forbid it. And
indeed, it was predicted theoretically~\cite{bhattacharjee,rondinelli} and observed
experimentally~\cite{tokura} that in the system $A$MnO$_3$ with magnetic Mn$^{4+}$
($d^3, S=\frac32$)  the system may become ferroelectric when one increases the size of
the $A^{2+}$ ions (going from Ca to Sr to Ba).

 But probably the more
natural route to create multiferroics in perovskites is the route first taken by
several groups (see reviews in~\cite{smolenskii,venevtsev}) --- to make a mixed
perovskite containing FE-active ions with the configuration~$d^0$, and
magnetic ions with configuration $d^n$, $n\neq0$, in the hope that every species will do ``what it
wants''. And indeed, on this route several multiferroics of this series were
synthesized, first by Smolenskii and his group~\cite{smolenskii}.
Some of them were even ferro-, or rather ferrimagnetic, which is very
favourable for possible applications. Unfortunately the coupling
between magnetic and FE degrees of freedom in these systems turned
out to be rather weak.

There are also other suggestions of how to make magnetic perovskites
ferroelectric. Thus, one idea is to use the coupling between rotation and
tilting of $B$O$_6$ octahedra often occurring in perovskites $AB$O$_3$~\cite{rondinelli-fennie}. And
indeed it was possible to create MF on this route. This
mechanism, however, belongs rather to another class of FE and MF
behaviour which we may call ``geometric'' mechanism.

\medskip

%<ESTOPPED>

\looseness=1
{\it ``Geometric'' multiferroics.} As mentioned in the footnote on p.~\pageref{fn:pfootn},
there are many multiferroics with improper FE, in which FE appears as a
secondary effect (a ``by-product'') of a primary ordering, e.g.\ of
rotation and tilting of structural units in a crystal, such as $M$O$_6$
octahedra. The best known and the most important examples of systems
with this mechanism of FE are the hexagonal systems $R$MnO$_3$ (where $R$ is a
small rare earth ion), e.g.\ YMnO$_3$. These systems sometimes are called
hexagonal perovskites, although they have not much in common with real
perovskites except similar-looking chemical formulae. In these
systems the Mn$^{3+}$ ions with 5-fold coordination are located in the centres
of trigonal bipyramids (two oxygen tetrahedra ``glued'' together by a
common face, common oxygen triangle). They form layered structures
with triangular Mn layers. Similar to conventional perovskites,
here these building blocks also have a tendency towards tilting and rotation,
to guarantee the close packing of the lattice. At such distortions (in
perovskites they are called GdFeO$_3$-distortion) there appear short $A$O
pairs with a dipole moment. But in perovskites $AB$O$_3$ these dipoles in
neighboring cells are oriented in opposite directions and cancel each
other, and in the standard case they don't lead to net ferroelectricity (unless
one uses special tricks to avoid this compensation). However in systems such as YMnO$_3$ there is no such
compensation. The geometric mechanism of ferroelectricity in these systems, which
presents a good example of improper ferroelectrics, was established in~\cite{palstra}.
These materials, with their interesting magnetic
structure, present nowadays an important playground for studying, in
particular, the characteristics of multiferroic domains and \hbox{domain walls}.

One extra comment must be made in connection with geometric ferroelectrics and
multiferroics. We have explained its origin by the rotation and tilting of building
blocks of a system, e.g.\ $M$O$_6$ octahedra in perovskites or $M$O$_5$ trigonal
bipyramids in systems such as YMnO$_3$. This one, in its turn, is usually
explained by the tendency towards close packing of rigid ions,
characterized, for perovskites $B$O$_3$, by the tolerance factor $t = (R_A + R_O)/\sqrt2(R_B+R_O)$:
for $t\sim1$ the system remains cubic, but for smaller values of~$t$
there occurs rotation and tilting of the $B$O$_6$ octahedra, see e.g.~\cite{khomskiiTM}.
But if we look more deeply, beyond the simplified picture of
rigid ions, we realize that it is again a certain tendency of
chemical bonds, in this case predominantly \hbox{$A$--O} bonds, which leads to
such distortions. Thus in effect it is always local chemistry
which is responsible for the formation and stability of one or the
other crystal structure, in particular FE one. Very often one can
express this tendency in the language of the pseudo-Jahn--Teller, or
second order Jahn--Teller effect~\cite{bersuker}.

\medskip

%\looseness=-1
{\it Lone pair mechanism.} Yet another ``chemical'' mechanism of ferroelectricity is
provided by materials containing ions with the so-called lone
pairs, or dangling bonds.  These are usually materials containing
Bi$^{3+}$ or Pb$^{2+}$. Bi typically accepts valences $3+$ and~$5+$. Bi$^{5+}$ has
the electronic structure (Xe)$4f^{14}5d^{10}$, %6s^26p^3$,
and Bi$^{3+}$ has two extra $6s$ electrons. In
principle these could become valence electrons and take
part in the formation of chemical bonds (as they do for Bi$^{5+}$).
However in Bi$^{3+}$, and similarly in Pb$^{2+}$,
% (Typical valencies of Pb  are Pb2+ )...6s^2) and Pb4+N(..., 6s^0)
these two electrons do not
participate in the formation of chemical bonds and are free to
``rotate'' in different directions in a crystal, which could lead to a
particular orientation of dipole moments associated with them. (Of
course these are not pure $6s$ electrons which are spherical, but they
are usually hybridized with their own $p$-electrons, or with
$p$-electrons of surrounding ligands, e.g.~oxygens.)

%The lone pair multiferroics are described in more details in Ch.~\#\# of this book;
%we will not dwell on it here.

\medskip

{\it Ferroelectricity due to charge ordering.} One more mechanism of FE
and of eventually MF behaviour is the possibility that the charge
ordering (CO), existing in some materials, can break inversion
symmetry and lead to ferroelectricity.
%This topic is discussed in great details in
%the chapter by Manuel Angst in this volume, thus
I will restrict
myself to a few remarks illustrating the main idea of this
phenomenon.

\begin{figure}[ht]
  \centering
  \includegraphics[scale=1]{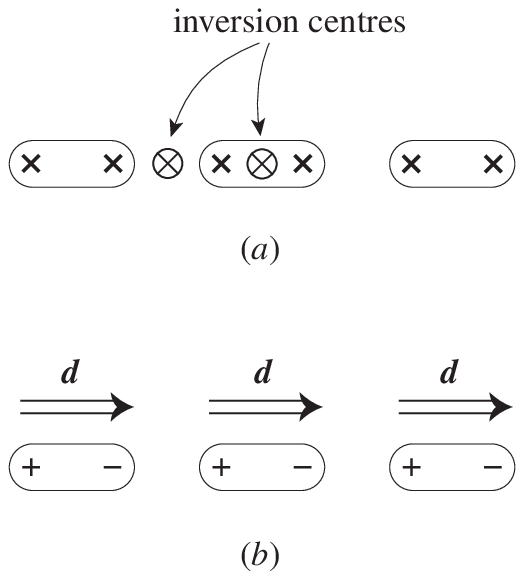}
  \caption{}
  \label{fig:7'}
\end{figure}

Suppose we have a structure consisting of dimers with equivalent
sites, such as for example H$_2$ molecules, see fig.~\ref{fig:7'}($a$).
This structure is definitely centrosymmetric and is not FE
(the inversion centres are marked in Fig.~\ref{fig:7'} by small encircled crosses~$\otimes$).
If however we now have an extra intradimer charge
ordering, Fig.~\ref{fig:7'}(b), making ``left'' and ``right'' ions in a dimer
inequivalent, each such dimer (a ``molecule'') would have a dipole
moment (double arrows in Fig.~\ref{fig:7'}(b)), and the entire system may become
ferroelectric. This mechanism was first proposed in~\cite{efremov}, and it is now
detected in some materials. And if
some constituting ions are magnetic, such systems would
simultaneously become multiferroic.

Concluding this section, a few general remarks are in order. When
considering electric polarization, one usually speaks about two
contribution to it, the ionic and the electronic contributions.
%Electronic mechanism of multiferroic behavior.  Before discussing
%specific mechanisms of purely electronic FE and MF behavior, one has
%to make some general comments.
FE transition is always a structural
phase transition from the paraelectric phase with centrosymmetric
crystal structure to the low-temperature state with lattice symmetry
with broken inversion, belonging to a pyroelectric class~\cite{landaulifshitz}.
As with all structural transitions, besides a shift
of ions there is also a redistribution of electron density (actually
a change of chemical bonding). Consequently, one can speak about two
contributions to the total polarization in ferroelectrics: an ion contribution and an
electronic one\footnote{Once again, if we look deeper, in effect all
cohesion in solids, and consequently all structural transitions, are
of electronic nature ``deep inside''. Nevertheless it makes sense to
separate electronic and ionic contribution to polarization --- with the
meaning, crudely speaking, that the ionic contribution is
predominantly the contribution of ionic cores, electronic
contribution being attributed to the change in the distribution of
valence electrons.} (although, strictly speaking, it may be
impossible to define those rigorously). By electronic contribution we
have in mind, more specifically, the FE caused by the change of
electronic distribution at fixed positions of ions. Of course, if we
then ``release'' the lattice, the ions would relax and the ionic
positions would adjust to the change of electronic density. Still,
the main driving mechanism could be predominantly electronic.

Usually these contributions are determined theoretically, for example
using {\it ab~initio} calculations. And the outcome turns out to be not
universal, and it strongly depends on the system. Thus, in perovskites
$R$MnO$_3$ with small rare earth $R = \rm Er$, \dots, with the $E$-type magnetic
structure, ionic and electronic contributions are comparable~\cite{picozzi},
but in TbMnO$_3$ the ionic contribution dominates, electronic
contribution being only about 10\% of the ionic one (and in opposite
direction)~\cite{vanderbild}. But in principle there can exist situations
with polarization of predominantly electronic character, see above
and the next section.

\section{Type-II multiferroics}
\label{sec:7}
By type-II multiferroics we refer to multiferroics in which ferroelectricity
exists only in a magnetically-ordered state, and is in fact {\it driven}
by a particular type of magnetic ordering.  It was the discovery of such MF~\cite{kimura,cheong}
that invigorated the entire field of multiferroics and which made them such
a hot topic.
From a symmetry point of view we are dealing here with materials in which
the crystal structure has an inversion symmetry, i.e.\ which in themselves
are not ferroelectric, but in which a particular type of {\it magnetic ordering}
breaks this inversion symmetry.  As discussed in Sec.~\ref{sec:4} above, if in such
situation both $\cal T$ and $\cal J$ invariance are broken, but the combined
$\cal JT$ invariance is preserved, the systems would be linear magnetoelectrics, but
not multiferroics.  If, however, the $\cal JT$ invariance is also broken, the system
is (or can become) multiferroic.

Microscopically one can also speak of several different groups of type-II multiferroics.
%These systems are described in detail in Ch.~\#\# of this book, thus
In this introductory
text
%chapter
I will give only a short overview of these
questions, paying main attention to some more general or more subtle points.

\medskip

\begin{figure}[ht]
  \centering
  \includegraphics[scale=0.9]{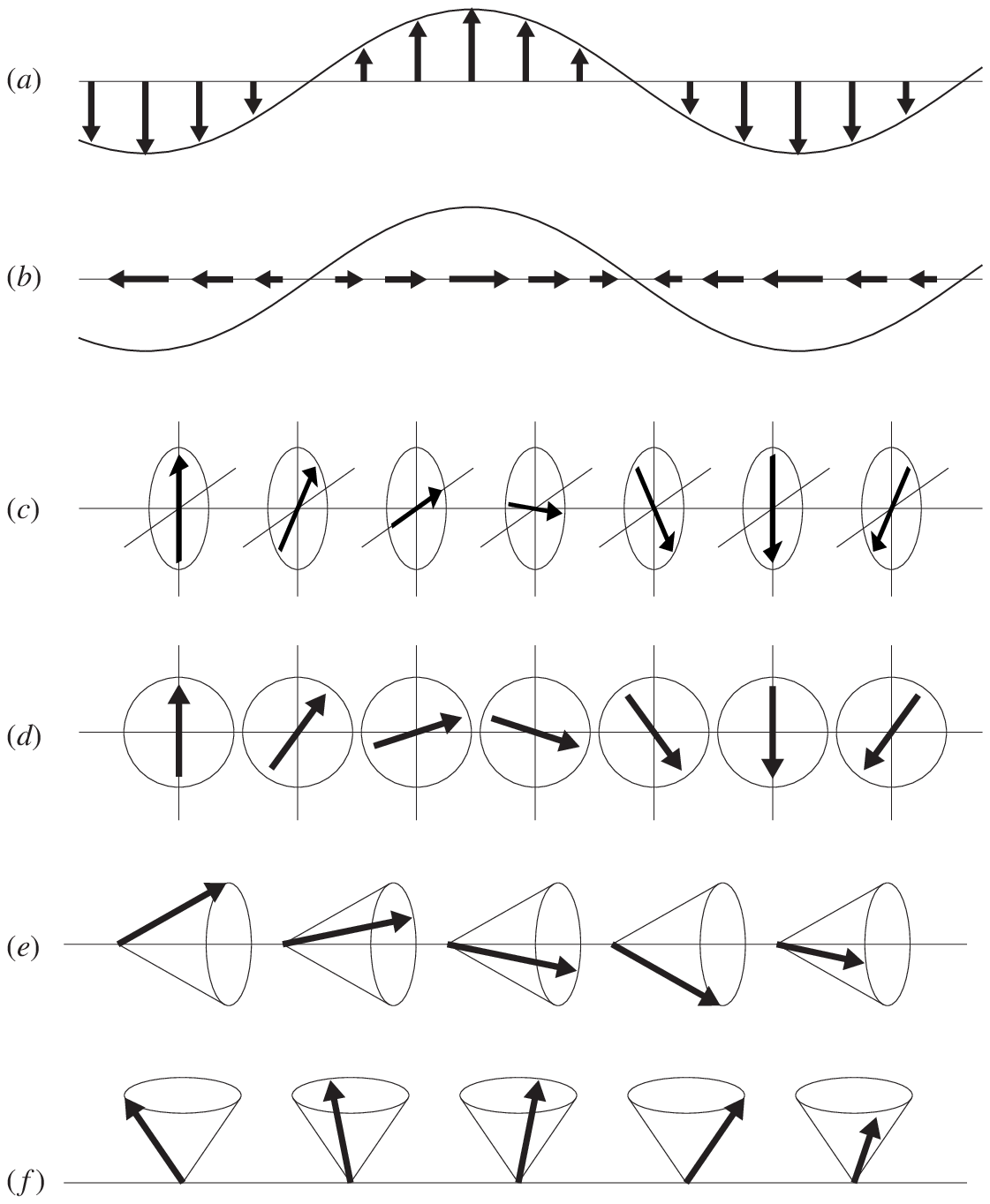}
  \caption{}
  \label{fig:8}
\end{figure}

{\it Type-II multiferroics with spiral magnetic structures.}  Probably
the big\-gest  group of type-II multiferroics discovered until now belong to
a class of materials with helicoidal or spiral magnetic structure.
Such structures are often incommensurate with the underlying crystal lattice, and
they present a subclass of spin density wave (SDW) structure.
We can have different types of SDW\null.
They can be sinusoidal, with spins perpendicular (Fig.~\ref{fig:8}($a$))
or parallel (Fig.~\ref{fig:8}($b$))
to the wavevector of the SDW\null.  Or they can be helicoidal (spiral) of two
types: proper screw, Fig.~\ref{fig:8}($c$), with spins rotating in the plane perpendicular to the wavevector,
or cycloidal, Fig.~\ref{fig:8}($d$), with the spin rotation plain containing the wavevector.
There may also exist different types of conical structures, two
of which are shown in Fig.~\ref{fig:8}($e$),~($f$).
As is shown both experimentally~\cite{kenzeli} and theoretically~\cite{katsura,dagotto,mostovoy},
in most cases ferroelectricity is produced by the cycloidal magnetic structures
of Fig.~\ref{fig:8}($d$).
Such are for example MF systems TbMnO$_3$, MnWO$_4$, etc.
The polarization of a pair of spins is given in this case by the following expression~\cite{katsura}:
\begin{equation}
\v P_{ij} \sim \v r_{ij} \times (\v S_i \times \v S_j)
\label{eq:8new}
\end{equation}
where $\v r_{ij}$ is the vector from site~$i$ to site~$j$, and $\v S_i$ and $\v S_j$
are the spins at corresponding sites.
For the cycloidal structure of Fig.~\ref{fig:8}($d$) the polarizations (dipole moments) of
consecutive bonds add, producing net ferroelectricity and multiferroic behaviour.
In this case one can write down the expression for the total polarization,~(\ref{eq:8new}),
in the equivalent form~\cite{mostovoy,cheong-mostovoy}
\begin{equation}
\v P \sim \v Q\times(\v S_i \times \v S_j) \sim \v Q\times \v e
\label{eq:9new}
\end{equation}
where $\v Q$ is the wavevector of the cycloid and $\v e$ is the axis
of rotation of spins in the cycloid.
In other words, the polarization of the cycloid lies in the plane
of rotating spins and perpendicular to the spiral axis.
The same expression also gives the polarization of the conical spiral of Fig.~\ref{fig:8}($f$):
the ``antiferromagnetic'' component of the spins rotates in a cycloidal fashion,
and produces polarization given by the same expressions (\ref{eq:8new}),~(\ref{eq:9new}).

According to these expressions, the proper screw of Fig.~\ref{fig:8}($c$) or conic spiral of Fig.~\ref{fig:8}($e$)
should not produce any polarization.  This was indeed the accepted point of view
for some time, until it was realized~\cite{arima, nagaosa2} that in certain cases these magnetic structures
can also lead to polarization, in this case directed {\it along} the spiral
direction~$\v Q$.
Indeed, in this case all directions perpendicular to~$\v Q$ are equivalent,
and polarization can be directed only along the spiral.
But if a crystal has two-fold symmetry axis $C_2$ perpendicular to~$\v Q$
(which was implicitly assumed in the derivation leading to expressions~(\ref{eq:8new}) and~(\ref{eq:9new})),
then both directions, parallel and antiparallel to~$\v Q$ are equivalent, i.e.\
$C_2\v P = -\v P$, which leads to $P=0$.
However if such two-fold axis in a crystal symmetry is absent,
nonzero polarization can in principle exist.  And indeed, several MF
materials
with the proper screw magnetic structure were discovered experimentally~\cite{CuMnO2}.

\begin{figure}[ht]
  \centering
  \includegraphics[scale=1]{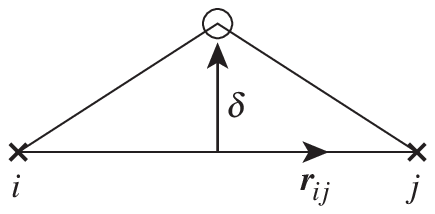}
  \caption{}
  \label{fig:9}
\end{figure}

From a microscopic point of view the mechanism of FE and MF in the most common
case of a cycloidal structure is the inverse Dzyaloshinskii--Moriya effect~\cite{dagotto}.
The Dzyaloshinskii--Moriya interaction has the form
\begin{equation}
\v D_{ij}\cdot(\v S_i \times \v S_j)\;,
\label{eq:9new-new}
\end{equation}
where the Dzyaloshinskii vector $\v D_{ij}$ for a pair~$ij$, in the case of
systems with superexchange e.g.\ via oxygen, Fig.~\ref{fig:9}, is proportional to the
displacement~$\delta$ of oxygen from the centre of the \hbox{$i$--$j$} bond,
\begin{equation}
\v D_{ij} \sim \v \delta \times \v r_{ij}\;.
\end{equation}
This interaction leads to the canting of otherwise collinear spins $\v S_i$,~$\v S_j$.
But, inversely, if by some reason (most often due to frustrations) the magnetic
structure is of a cycloidal type with canted neighbouring spins, it is favourable
to shift the oxygens to gain the Dzyaloshinskii--Moriya energy~(\ref{eq:9new-new}).
For a cycloidal structure  of Fig.~\ref{fig:8}($d$) all such shifts would be in the same direction,
which would produce an electric polarization given by the expressions (\ref{eq:8new}),~(\ref{eq:9new}).

There are also other microscopic mechanisms leading to MF in such
cases, e.g.\ for a proper screw~\cite{nagaosa2}.  I will not discuss them here.
It is only important to notice that all of them rely on the presence
of the (relativistic) spin--orbit interaction $\lambda\v l\cdot \v S$,
and therefore typically the FE polarization in such cases is rather small.
There exists however another mechanism of multiferroicity, not relying
on the real spin--orbit interaction and acting also for collinear magnetic structures.

\medskip

{\it Magnetostriction mechanism of type-II multiferroics.}  Another mechanism
of multiferroic behaviour
%, described in more detail in Ch.~\#\#,
is the
standard magnetostriction due to the dependence of the exchange integral~$J_{ij}$
on the distance between sites $i$ and~$j$, and often also on the angle
\hbox{$M_i$--O--$M_j$} for the superexchange mechanism, Fig.~\ref{fig:9}.
Due to this magnetostriction some ions shift in the magnetically-ordered
phase, which can lead to electric polarization.

\begin{figure}[ht]
  \centering
  \null\qquad\qquad\includegraphics[scale=1]{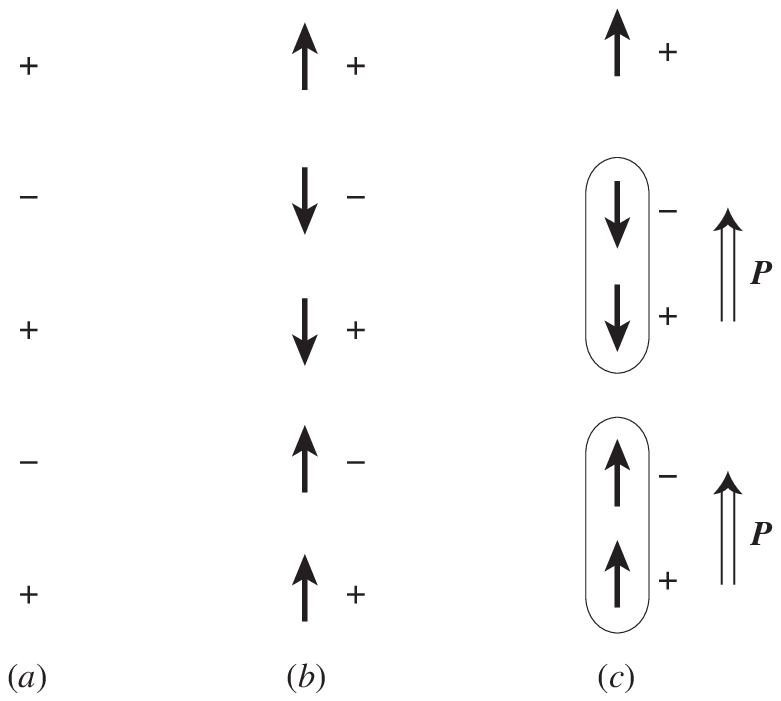}
  \caption{}
  \label{fig:10}
\end{figure}

The simplest (and actually realistic) example is shown in Fig.~\ref{fig:10} (cf.\
Fig.~\ref{fig:7'}, with which it has much in common, and also the discussion at the
end of Sec.~4).  Suppose we have a lattice made of ions with different
charges, which in itself is centrosymmetric, Fig.~\ref{fig:10}($a$) (each ion is
an inversion centre).  If however the magnetic structure is of the type
${\uparrow}\,{\uparrow}\,{\downarrow}\,{\downarrow}$, Fig.~\ref{fig:10}($b$), then the
inversion symmetry would be broken by this magnetic structure, and
the material could become ferroelectric. And indeed, due to magnetostriction
the ferromagnetic and antiferromagnetic bonds would become inequivalent,
and if, for example, the ferromagnetic bond would become shorter, we
would get the situation of Fig.~\ref{fig:10}($c$), with polarization shown by the double arrow,~$\Uparrow$.
We see that this mechanism resembles very much that of Fig.~\ref{fig:7'},
where we started with inequivalent bonds and obtained ferroelectricity
by putting charge ordering on top.  Here we consider an opposite situation:
with inequivalent sites, bonds becoming inequivalent due to magnetostriction
in a particular magnetic structure.  Note that this
mechanism works also for a collinear spin structure, and it does not
require any relativistic spin--orbit interaction.  Therefore one could
expect larger values of FE polarization in such systems, and indeed
this is the case: if for typical cycloidal multiferroics the polarization
is usually $\sim 10^{-2}\,\rm \mu C/cm^2$, in multiferroics with
the magnetostriction mechanism it can reach several~$\mu C/cm^2$.
Thus, theoretical considerations predict for Mn perovskites
with $E$-type magnetic structure (resembling somewhat that of Fig.~\ref{fig:10}($b$))
the polarization~$\sim2\,\rm \mu C/cm^2$~\cite{picozzi}.
The first measurements~\cite{lorenz} demonstrated much smaller values, but
the improved quality of the samples allowed to reach values almost
equal to the theoretical prediction.

Experimentally there exist many multiferroics with this mechanism
of ferroelectricity: e.g.\ one of the very first type-II multiferroics,
TbMn$_2$O$_5$~\cite{cheong,chapon}; the material Ca$_3$CoMnO$_6$~\cite{choi}
which is well described by the structure of Fig.~\ref{fig:10}; CdV$_2$O$_4$~\cite{giovanetti};
and a few others.

\medskip

\begin{figure}[ht]
  \centering
  \includegraphics[scale=1]{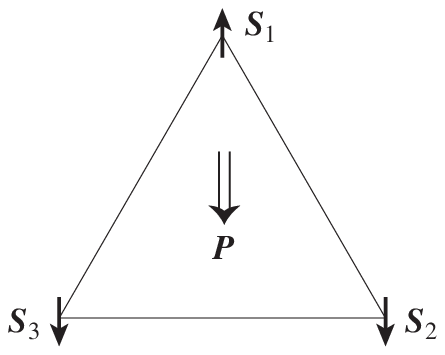}
  \caption{}
  \label{fig:10'}
\end{figure}

{\it Electronic mechanism of ferroelectricity in frustrated magnets.}
If in the examples of type-II multiferroics considered above the ionic
displacements played a crucial role, and the ionic contribution
to polarization was significant, sometimes dominant~\cite{vanderbild}
(although the electronic contribution was also important), there
exists a possibility of a purely electronic mechanism of MF behaviour.
Such mechanism, proposed in~\cite{ilexi}, see also~\cite{khomskii-extra}, can operate
in frustrated magnets with a particular magnetic structure.
One can show that in a magnetic triangle with one electron per site,
described by the strongly interacting Hubbard model
\begin{equation}
{\cal H} = -t\sum_{\langle ij\rangle}c^\dagger_{i\sigma}c\fd_{j\sigma} +  U\sum_i n_{i\uparrow}n_{i\downarrow}
\end{equation}
there would appear, for certain magnetic structures, a charge redistribution, given
by the expression
\begin{equation}
n_i = 1 + \frac{32t^3}{U^2}\Bigl[\v S_1\cdot(\v S_2 + \v S_3)  -  2\v S_2\cdot\v S_3\Bigr]\;,
\label{eq:12new}
\end{equation}
see Fig.~\ref{fig:10'}.  If the spin correlation function entering (\ref{eq:12new}) is nonzero,
as is the case for the structure of Fig.~\ref{fig:10'}, charge redistribution would occur,
and the triangle would acquire a dipole moment (double arrow in Fig.~\ref{fig:10'}).
In principle this could give net polarization in a bulk solid consisting of such triangles,
see~\cite{khomskii-extra}, in which case the material would be multiferroic.
(Of course, if we then release the lattice, it would also distort somewhat, also
contributing to the polarization; but the main mechanism and the driving force
of multiferroic behaviour is in this case purely electronic.)

\section{Beyond multiferroics}
\label{sec:8}
It is a very interesting development that, after learning many things
while studying multiferroics, we can go back to some more traditional magnetic
systems, and using the ``multiferroic know-how'' predict (and observe) many very
nontrivial effects connected with the coupling between magnetic and electric phenomena.
I will briefly discuss a few of these below.
%; a more detailed treatment can be found in other chapters of this book.

\subsection{Electric activity of magnetic domain walls}
The specific features of domain wall in multiferroics such as BiFeO$_3$ or YMnO$_3$
is a very important and well-studied field.
Here I want to draw attention to a slightly different aspect.  It was first pointed out
by Mostovoy~\cite{mostovoy} that there should exist nontrivial magnetoelectric
and multiferroic effects in some domain wall even in ordinary insulating ferromagnets.

\begin{figure}[ht]
  \centering
  \includegraphics[scale=0.9]{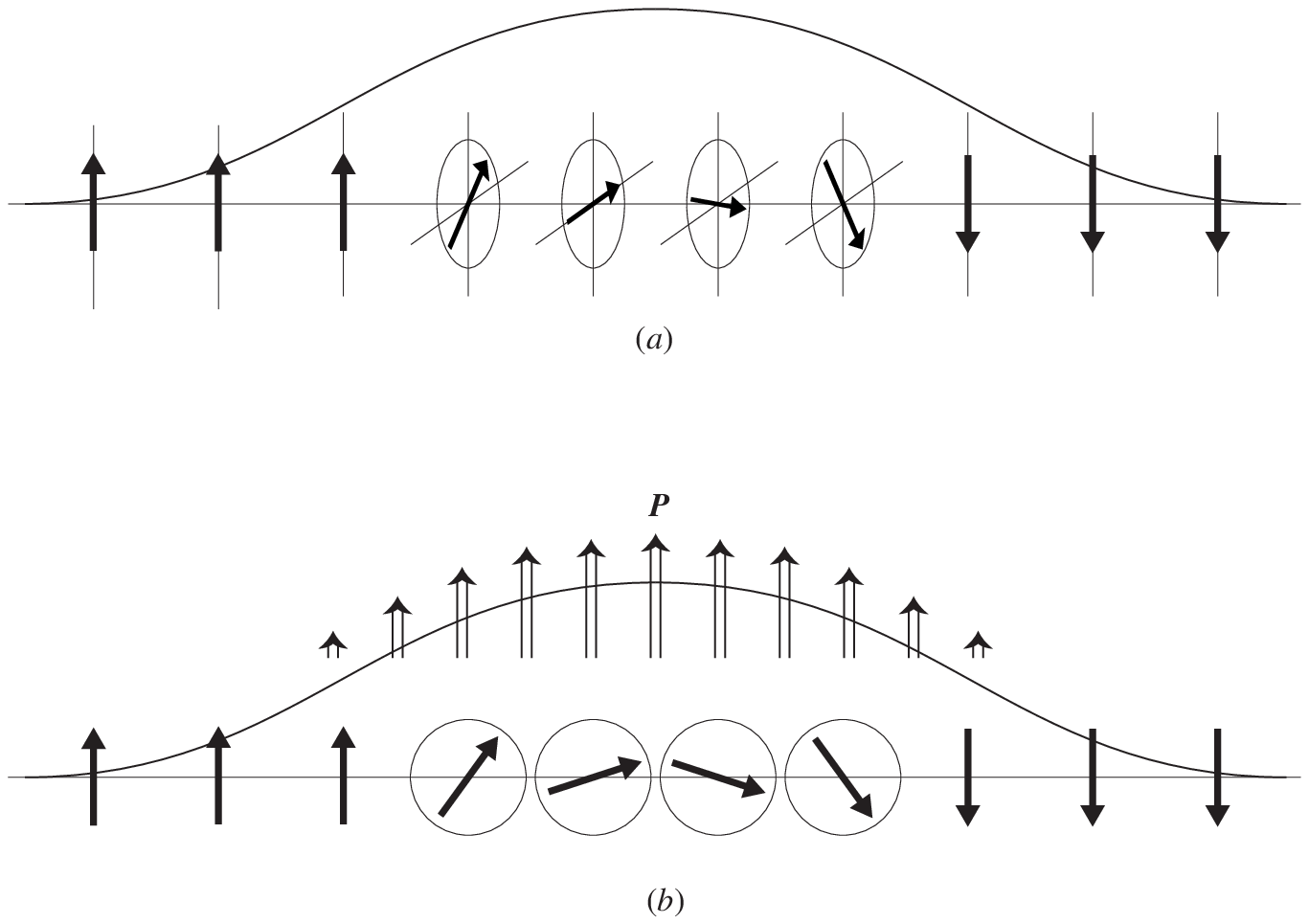}
  \caption{}
  \label{fig:11}
\end{figure}

Typically, there exist two types of domain walls in ferromagnets: Bloch domain walls,
Fig.~\ref{fig:11}($a$), in which spins rotate in the plane of the wall, perpendicular to direction from one
domain to the other; and N\'eel domain wall, Fig.~\ref{fig:11}($b$), in which spins in the centre of the domain wall
point alongside the normal to this wall.
We immediately see that the Bloch wall presents a part of the proper screw of Fig.~\ref{fig:8}($c$),
whereas the N\'eel wall has a ``cycloidal'' structure of Fig.~\ref{fig:8}($d$).
According to eqs.~(\ref{eq:8new}),~(\ref{eq:9new}) one should then expect
the appearance of electric polarization at every N\'eel domain wall.

One can then propose a beautiful experiment, which was indeed carried out~\cite{logginor}.
It is well known that if we put magnetic dipoles (``magnetic needles'') in an
inhomogeneous magnetic field, such needles would be attracted (or repelled, depending
on the direction of the dipoles) to the region of stronger field.  The same
of course is also true for electric dipoles in the gradient of an electric field.

\begin{figure}[ht]
  \centering
  \includegraphics[scale=1]{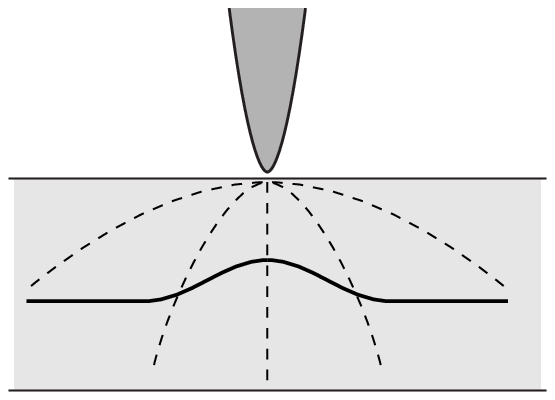}
  \caption{}
  \label{fig:12}
\end{figure}

The group of Logginov and Pyatakov~\cite{logginor} carried out such experiment
not with the usual magnetic dipoles, but with the insulating ferromagnets
with N\'eel domain walls, which, according to arguments presented above,
should carry electric dipoles.  They used a film of a well-known such material,
the iron garnet (Bi,Lu)$_3$(Fe,Ga)$_5$O$_{12}$, with $T_{\rm c}\sim450\,\rm K$, approached it by a sharpened
copper wire and applied a voltage pulse to the wire.
This produced an inhomogeneous electric field in the film, see Fig.~\ref{fig:12},
and it was observed that the N\'eel domain walls were attracted
to the region of stronger electric field.
This experiment, besides demonstrating the appearance of electric dipoles
on N\'eel domain walls, opens a way to control such domain wall by the electric field,
which may be extremely useful in manufacturing new electrically-controlled memory media and devices.

\subsection{Spiral magnetic structures on metal surfaces}
\begin{figure}[ht]
  \centering
  \includegraphics[scale=1]{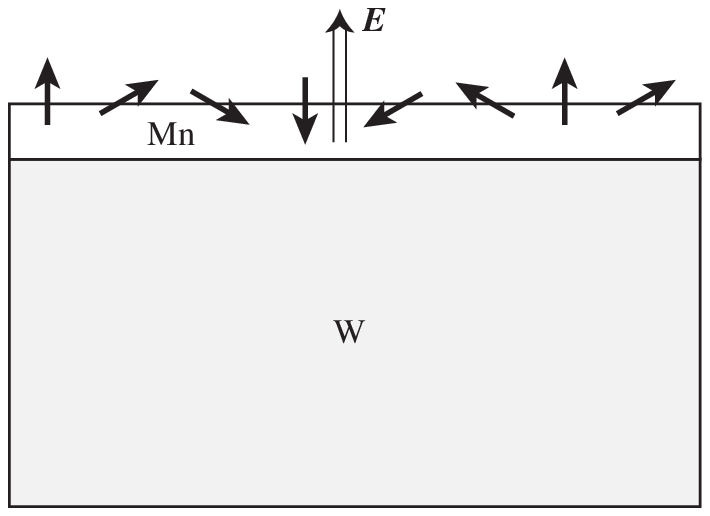}
  \caption{}
  \label{fig:13}
\end{figure}

Yet another experimental observation which can be easily explained if we invoke
the physics described above is the detection of helicoidal magnetic
structures in thin films (monolayer, bilayer) of magnetic metals on nonmagnetic
substrates~\cite{bode,kubetska,heide}.
The first such results were obtained on a monolayer of manganese on tungsten~\cite{bode}, see Fig.~\ref{fig:13}.
It was observed that instead of forming a collinear magnetic structure, a cycloidal
spiral was formed (Fig.~\ref{fig:13} is a simplified schematic picture, which
shows the situation if the Mn layer were ferromagnetic; see the original publication for the actual
structure).  From what we have learned in Sec.~\ref{sec:7},  it is clear
 that the cycloidal spiral would produce a polarization and a corresponding electric
field, perpendicular to the surface.  But, vice versa, the intrinsic
electric field always existing at the surface of a metal (due to the double-charge
layer, or the potential drop --- the work function of the metal), which is normal to the
surface, would have the tendency to create a cycloidal spiral from the initially collinear
spin structure.  And as this intrinsic electric field in the double layer always has the same sign
on the entire surface, all the spirals in such monolayer would have the same
sense of rotation, e.g.\ clockwise.

This phenomenon should be quite general,
since such potential drop exists at the surface of every metal.
But this tendency to rotate the spins would act against the magnetic anisotropy,
always present at the surface, and in order to be observable it must be strong
enough to overcome this anisotropy. This is why  such cycloidal
structures are found not on all  magnetic layers on top of any metal.
One needs at least a strong spin--orbit coupling at the surface
(we remind that the mechanism leading to (\ref{eq:8new}),~(\ref{eq:9new}), is in fact the spin--orbit interaction),
for which it is better to use a heavy metal as a substrate; and it is also
desirable to have magnetic metals without very strong single-site anisotropy.

The authors of the original papers themselves proposed an explanation which relies
on the Dzyaloshinskii--Moriya interaction at the surface~\cite{heide}
(which always exists in this case, since the inversion symmetry is broken
by the surface itself).  In fact, physically this is the same explanation
as the one described above (as mentioned previously, the microscopic mechanism
of the relations (\ref{eq:8new}),~(\ref{eq:9new}) is in fact
the same Dzyaloshinskii--Moriya interaction).
However the picture presented in Fig.~\ref{fig:13} is simpler and more transparent, even
if it is conceptually the same.  By the way, the same coupling of magnetic
and electric degrees of freedom at the surface, which lies at the core
of our picture, is also very similar to the recent proposal~\cite{kim}
explaining the large magnitude of Rashba spin--orbit coupling at the surface
by the important role of electric dipoles at the surface layer.

\subsection{Magnetoelectric effects on magnetic vortices and skyrmions}
\begin{figure}[ht]
  \centering
  \includegraphics[scale=1]{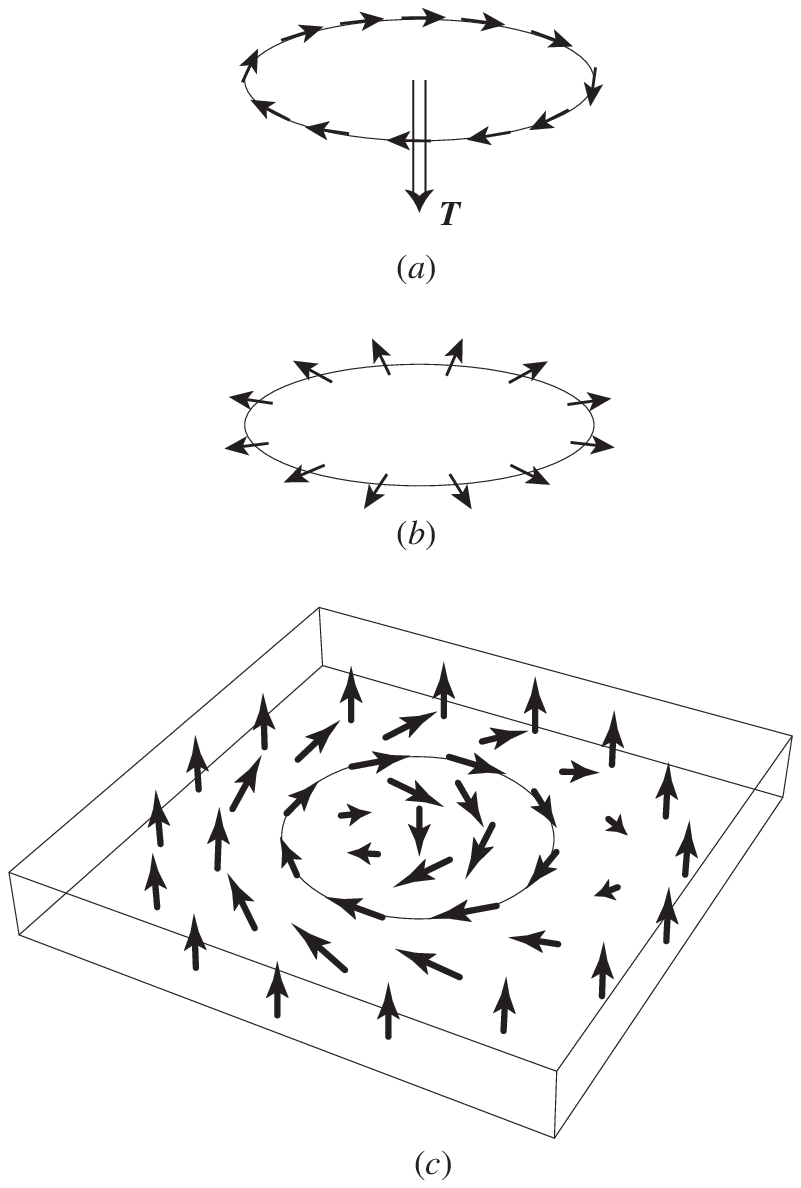}\quad\null
  \caption{}
  \label{fig:14}
\end{figure}

There may exist different magnetic textures in normal magnets: not only
domain walls, but also, for example, magnetic vortices, Fig.~\ref{fig:14}($a$),~($b$),
or skyrmions, Fig.~\ref{fig:14}($c$).

One can show that all such objects would exhibit a linear magnetoelectric effect~\cite{montory}.
The ``head to tail'' vortices of Fig.~\ref{fig:14}($a$) are classical examples of
systems with toroidal moment~$\v T\sim \sum_i\v r_i\times \v S_i$, shown
in Fig.~\ref{fig:14}($a$) by a double arrow.
This means that here we would have a ME of the type described in Sec.~\ref{sec:6},
with e.g.\ the polarization~$\v P$ perpendicular to the magnetic field~$\v H$,
$\v P \sim \v T\times\v H$, and similarly with magnetization~$\v M$ induced by
electric field~$\v E$, $\v M\sim\v T\times\v E$.
On the other hand, the ``radial'' vortex of the type of Fig.~\ref{fig:14}($b$) would have
a linear ME effect with a {\it symmetric} ME tensor $\alpha_{ij}$, i.e.\
the polarization would be parallel to the external magnetic field
(along the principal directions in which $\alpha_{ij}$ is diagonal).

Skyrmions (``magnetic hedgehogs'') can be of two types.  Most often, Fig.~\ref{fig:14}($c$),
in the ``middle part'' of a skyrmion
the spins are rotating as in Fig.~\ref{fig:14}($a$).  In this case the skyrmions would
also exhibit transverse ME effect.  But in the case of ``radial'' skyrmions
(real ``hedgehogs'') the ME would be longitudinal.  The ME effect
in skyrmions was recently experimentally observed in~\cite{okamura-skyrm}.
Interestingly enough, skyrmions can also be created on magnetic layers
on the surface of a metal~\cite{heinze,romming}, apparently due to the mechanism described in the
previous subsection.

\subsection{Electric activity of spin waves}
\begin{figure}[ht]
  \centering
  \includegraphics[scale=1]{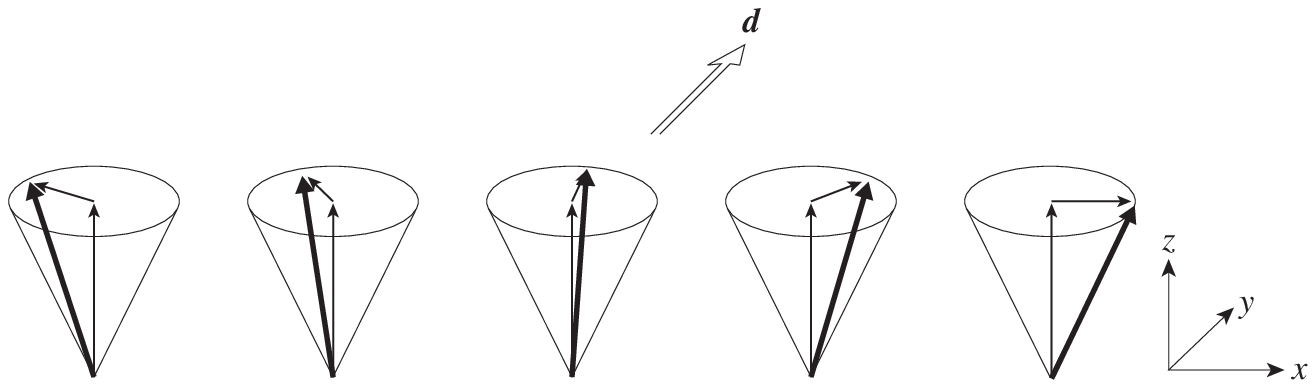}
  \caption{}
  \label{fig:15}
\end{figure}

From what we have learned in studying the cycloidal type-II multiferroics, Sec.~\ref{sec:7},
we can predict yet another interesting effect --- electric activity of spin waves~\cite{khomskii-trends}.
As is well known, see e.g.~\cite{khomskiiTM}, a quasiclassical picture of
magnons in a ferromagnet is the precession of spins along the direction
of magnetization, Fig.~\ref{fig:15}.  As we see, in this case the instantaneous
picture (a ``snapshot'') is that in which the perpendicular ($xy$) component of magnetization
forms a cycloid in $xy$-plane.  Consequently, according to our understanding
reached in studying type-II multiferroics, eqs.\ (\ref{eq:8new}),~(\ref{eq:9new}),
such spin wave should carry, besides magnetization, also an electric dipole
perpendicular both to magnetization and to the magnon wavevector~$\v Q$ (double
arrow in Fig.~\ref{fig:15}).

\begin{figure}[ht]
  \centering
  \includegraphics[scale=1]{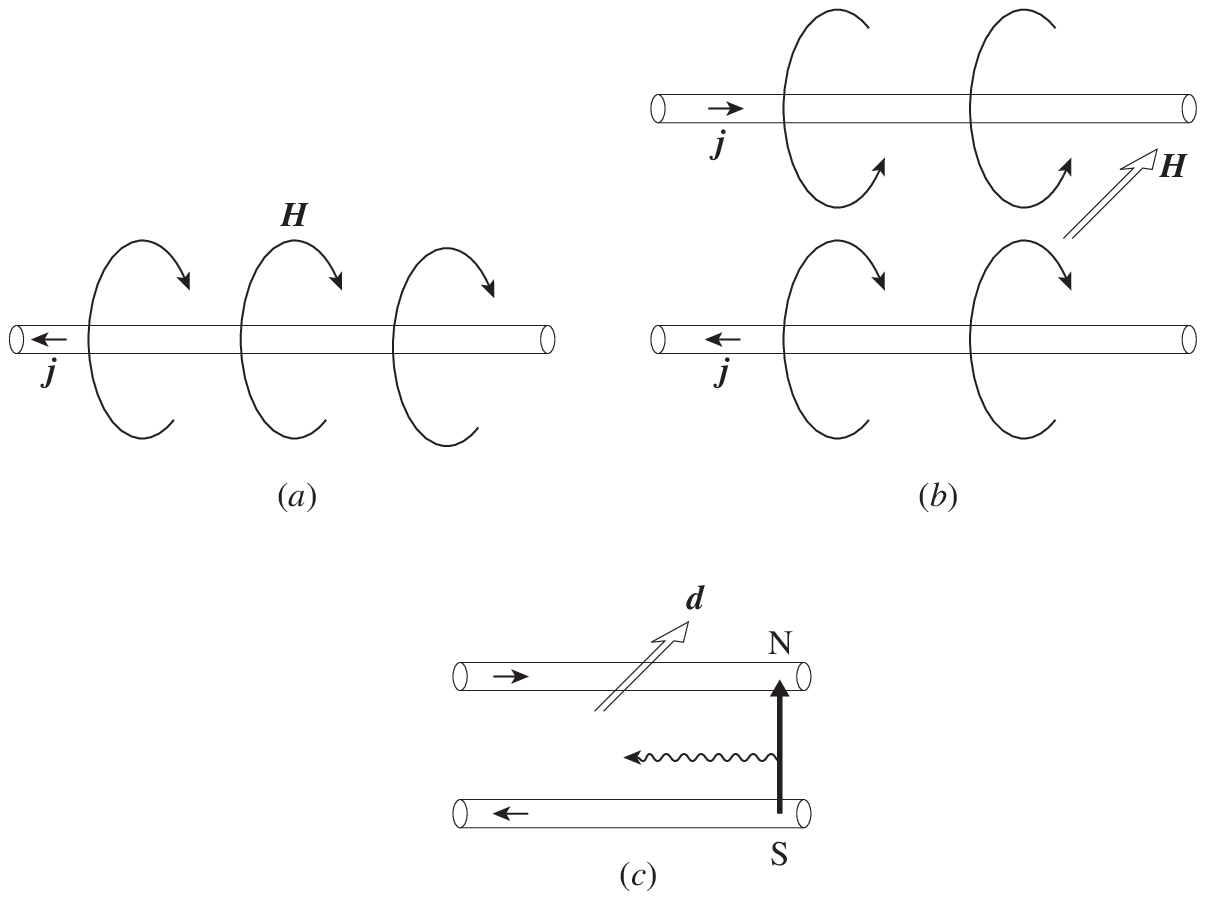}
  \caption{}
  \label{fig:16}
\end{figure}

One can also give this effect a completely classical interpretation.
It is well-known that there exists a circular magnetic field around a wire
carrying a current, Fig.~\ref{fig:16}($a$).  Correspondingly, the field of two such
wires with currents in opposite directions would produce a magnetic field,
or magnetic moment, located between the wires and pointing in the direction
marked on Fig.~\ref{fig:16} by a double arrow. (The fields far away from the double
wire would cancel.)  But there exists a well-known symmetry between
magnetic and electric phenomena, exemplified best of all by the Maxwell equations.
Correspondingly, the motion of a magnetic dipole, which can be represented
as a parallel motion of south and north poles, or positive and negative magnetic charges
(although such monopoles do not exist separately), would correspond to
magnetic currents running in opposite directions, as in Fig.~\ref{fig:16}($b$), and in effect
such ``currents'' created by a magnon --- classically the propagating magnetic
dipole --- would be accompanied by an electric field, or electric
dipole shown in Figs.~\ref{fig:15},~\ref{fig:16}($c$).

\section{Conclusions}
In this introductory
text
%chapter
I have tried to give an overview of some basic
notions and phenomena which we meet in studying this interesting class
of materials --- multiferroics.
%A more detailed exposition of particular topics will be given in the next chapters of this book.
Concluding this short overview I only want to stress once again
that not only multiferroics themselves present significant interest,
both from general point of view and that of practical applications,
but also the study of other interesting phenomena in other types
of solids benefits strongly from the experience we get in studying
multiferroics.


\begin{thebibliography}{99}

\raggedright

\bibitem{fiebig}M.~Fiebig,
%{\it Revival of the magnetoelectric effect},
J. Phys. D: Appl. Phys. {\bf 38}, R123

\bibitem{cheong-mostovoy}S.-W. Cheong, M.~Mostovoy,
%{\it Multiferroics: a magnetic twist for ferroelectricity},
Nature Materials 6: 13--20

\bibitem{khomskiiJMMM}D.~I.~Khomskii,
%{\it Multiferroics: Different ways to combine magnetism and ferroelectricity,}
J. Mag. Mag. Mat. 306: 1

\bibitem{ehrenreich}W.~Eerenstein, N.~D.~Mathur, J.~F.~Scott,
%{\it Multiferroic and magnetoelectric materials},
Nature 442: 759

\bibitem{kitaicy}K.~F.~Wang, J.~M.~Liu, Z.~F.~Ren,
%{\it Multiferroicity: the coupling between magnetic and polarization orders}.
Adv. Phys. 58: 321--448

\bibitem{nagaosa}Y.~Tokura,
%{\it Multiferroics --- toward strong coupling between magnetization and polarization in a solid},
J. Mag. Mag. Mat. 310: 1145--1150

\bibitem{khomskii-trends}D.~I.~Khomskii,
%{\it Multiferroics: mechanisms and effects},
Physics (Trends) {\bf 2}, 20 (2009)

\bibitem{JPCM}Special issue, J. Phys. Condens. Matter 20, 434201--434220 (2008).

\bibitem{khomskiiTM}D.~I.~Khomksii, {\it Transition Metal Compounds}, Cambridge
University Press, Cambridge 2014

\bibitem{curie}P.~Curie, J. Phys. (Paris) Colloq. 3, 393 (1894)

\bibitem{landaulifshitz}L.~D.~Landau and E.~M.~Lifshitz, {\it Electrodynamics of Contintuous Media},
Moscow 1959; English edition: Pergamon Press, Oxford 1960

\bibitem{dzyaloshinskii}I.~E.~Dzyaloshinskii, Sov. Phys. JETP 10, 628 (1959)

\bibitem{astrov}D.~N.~Astrov, Sov. Phys. JETP 11, 708 (1960)

\bibitem{freeman}A.~J.~Freeman and H.~Schmid, eds.,
{\it Magnetoelectric Interaction Phenomena in Crystals}, Gordon and Breach, London, New York, Paris 1975

\bibitem{schmid-boracite}E.~Ascher, H.~Rieder, H.~Schmid, and H.~S\"ossel, J. Appl. Phys. {\bf 37}, 1404 (1966)

\bibitem{schmid-term}H.~Schmid, Ferroelectrics {\bf 162}, 317 (1994)

\bibitem{goodenough-longo}J.~B.~Goodenough and J.~M.~Longo, {\it Magnetic and Other Properties
of Oxides and Related Compounds}, Landolt-B\"ornstein, {\it Numerical
Data and Functional Relations in Science and Technology}, New
Series Vol. III. 4 (Springer, Berlin, 1970)

\bibitem{landolt}T.~Mitsui {\it et al.}, {\it Ferroelectrics and Related Substances},
Landolt-B\"ornstein, {\it Numerical data and Functional Relations in Science
and Technology}, New Series Vol.~16 (1) (Springer, Berlin, 1981)

\bibitem{APS}D.~I.~Khomskii, Bull. Am. Phys. Soc. C 21.002 (2001)

\bibitem{hill}N.~A.~Hill, J. Phys. Chem. B 104, 6694 (2000)

\bibitem{wang}J.~Wang {\it et al.}, Science {\bf 299}, 1719 (2003).

\bibitem{kimura}T.~Kimura {\it et al.}, Nature {\bf 426}, 55 (2003)

\bibitem{cheong}N.~Hur {\it et al.}, Nature {\bf 429}, 392 (2004).

\bibitem{zwezdin}A.~K.~Zvezdin {\it et al.}, JETP Lett. {\bf 83}, 509 (2006)

\bibitem{strickman}W.~F.~Brown, R.~M.~Hornreich, S.~Shtrickman,
Phys. Rev. {\bf 168,} 574 (1968)

\bibitem{electromagnons}A.~Pimenov {\it et al.}, Nature Phys. {\bf 2}, 97 (2006)

\bibitem{spaldin}N.~A.~Spaldin, Journal of Solid State Chemistry,
{\bf 195}, 2 (2012)

\bibitem{schmid-review}H.~Schmid, J. Phys. Condens. Matter {\bf 20}, 434201 (2008)

\bibitem{choi}Y.~J.~Choi {\it et al.}, Phys. Rev. Lett. {\bf 100}, 047601 (2008)

\bibitem{vdBrink}J.~van den Brink and D.~I.~Khomskii,
%{\it Multiferroicity due to charge ordering},
J. Phys. Cond. Matt. {\bf 20}, 434217 (2008)

\bibitem{PbVO3-1}R.~V.~Shpanchenko {\it et al.}, Chem. Mater. {\bf 16}, 3267 (2004)

\bibitem{PbVO3-2}A.~A.~Belik, M.~Azuma, T.~Saito {\it et al.},
Chem. Mat. {\bf 17}, 2 269 (2005)

\bibitem{aharony}A.~B.~Harris, A.~Aharony, O.~Entin-Wohlman,
J. Phys. Cond. Matt. {\bf 20}, 43 434202 (2008)

\bibitem{viret}D.~Lebeugle {\it et al.},
Phys. Rev. Lett. {\bf 100}, 227602 (2008)

\bibitem{levanyuk}A.~P.~Levanyuk and D.~G.~Sannikov, Sov. Phys.--Uspekhi 17, 199 (1974).

\bibitem{LLstat}L.~D.~Landau and E.~M.~Lifshitz, {\it Statistical Physics}, Addison-Weslsy, Reading,
Mass., 1969; or 3d ed. Part 1, Butterworth-Heinemann, 1980

\bibitem{khomskii-aspects}D.~I.~Khomskii, {\it Basic Aspects of Quantum Theory of
Solids: Order and Elementary Excitations}, Cambridge University Press, Cambridge 2010

\bibitem{Shannon}R.~D.~Shannon, {Acta Crystallographica} {\bf 751} (1975)

\bibitem{abrikosov}A.~A.~Abrikosov, L.~P.~Gorkov, Sov.~Phys.--JETP {\bf 12}, 1242 (1961)

\bibitem{bhattacharjee} S.~Bhattacharjee, E.~Bousquet, and P.~Ghozes, Phys. Rev.
Lett. {\bf 102}, 117602 (2009)

\bibitem{rondinelli}J.~M.~Rondinelli, A.~S.~Eidelson, and N.~A.~Spaldin, Phys.
Rev. B {\bf 79}, 205119 (2009)

\bibitem{tokura}H.~Sakai, J.~Fujioka, T.~Fukuda {\it et al.},
Phys. Rev. Lett. {\bf 107} 137601 (2011)

\bibitem{smolenskii}G.~A.~Smolenskii {\it et al.}, {\it Segnetoelectrics and Antisegnetoelectrics},
Nauka Publishers, Leningrad (1971);
G.~A.~Smolenskii and I.~E.~Chupis, Sov. Phys. Usp. {\bf 25}, 475 (1982)

\bibitem{venevtsev}Y.~N.~Venevtsev, V.~V.~Gagulin, Ferroelectrics {\bf 162}, 23 (1994)

\bibitem{rondinelli-fennie}J.~M.~Rondinelli and C.~J.~Fennie, Adv.~Mayer. {\bf 11}, 354 (2010)

\bibitem{palstra}B.~van~Aken, T.~T.~M.~Palstra, A.~Filippetti, N.~A.~Spaldin, Nature
Mater. {\bf 3} (2004) 164

\bibitem{bersuker}I.~B.~Bersuker, {\it The Jahn--Teller Effect}, Cambridge Univ. Press, Cambridge 2006

\bibitem{efremov}D.~V.~Efremov, J.~van~den~Brink, and D.~I.~Khomskii, Nature Mater. {\bf 3}, 853 (2004)

\bibitem{picozzi}S.~Picozzi {\it et al.}, Phys. Rev. B {\bf 74}, 094402 (2006);
%A.~Malashevich and D.~Vanderbilt, Phys. Rev. Lett. {\bf 101}, 037210 (2008)

\bibitem{vanderbild}A.~Malashevich and D.~Vanderbild, Phys. Rev. Lett. {\bf 101}, 037210 (2008)

\bibitem{kenzeli}M.~Kenzelmann {\it et al.}, Phys. Rev. Lett. {\bf 98}, 267205 (2007).

\bibitem{katsura}H.~Katsura, N.~Nagaosa, and A.~V.~Balatsky, Phys. Rev. Lett. {\bf 95},
057205 (2005)

\bibitem{dagotto}I.~A.~Sergienko, C.~Sen, and E.~Dagotto, Phys. Rev. Lett. {\bf 97}, 227204
(2006).

\bibitem{mostovoy}M.~V.~Mostovoy, Phys. Rev. Lett. {\bf 96}, 067601 (2006)

\bibitem{arima}T.~Arima, J. Phys. Soc. Jpn. {\bf 76}, 073702 (2007)

\bibitem{nagaosa2}C.~Jia, S.~Onoda and N.~Nagaosa,
Phys. Rev. B. {\bf 76} 1444424 (2007)

\bibitem{CuMnO2}T.~Kimura, J.~C.~Lashley, A.~P.~Ramirez, Phys. Rev. B {\bf 73}, 220401
(2006); S.~Seki {\it et al.}, Phys. Rev. Lett. {\bf 101}, 067204 (2008); K.~Kimura
{\it et al.}, Phys. Rev. B {\bf 78}, 140401 (2008)

\bibitem{lorenz}B.~Lorenz, Y.~Q.~Wang, and C.~W.~Chu, Phys. Rev. B {\bf 76}, 104405
(2007)

\bibitem{chapon}P.~G.~Radaelli, L.~C.~Chapon,
J.~Phys.~Cond.~Matt. {\bf 20} 43, 434213 (2008)

\bibitem{giovanetti}G.~Giovanetti {\it et al.}, Phys. Rev. B {\bf83}, 060402 (2001)

\bibitem{ilexi}L.~N.~Bulaevskii {\it et al.}, Phys. Rev. B {\bf 78}, 024402 (2008)

\bibitem{khomskii-extra}D.~I.~Khomskii,
%``Spin chirality and nontrivial charge dynamics in frustrated Mott insulators: spontaneous currents and
%charge redistribution'',
Journal of Physics -- Condensed Matter {\bf 22}, 164209 (2010)

\bibitem{logginor}A.~S.~Logginov {\it et al.}, JETP Lett. {\bf 86}, 115 (2007);
Appl. Phys. Lett. {\bf 93}, 182510 (2008)

\bibitem{bode}M.~Bode {\it et al.}, Nature {\bf 447}, 190 (2007); P.~Ferriani {\it et al.}, Phys. Rev.
Lett. {\bf 101}, 027201 (2008)

\bibitem{kubetska}A.~Kubetska {\it et al.}, Phys. Rev. Lett. {\bf 88}, 057201 (2002); E.~Y.~Vedmedenko
{\it et al.}, Phys. Rev. Lett. {\bf 92}, 077207 (2004)

\bibitem{heide}M.~Heide, G.~Bihlmayer and S.~Bluegel, Phys. Rev. B {\bf 78}, 140403 (2008)

\bibitem{kim}S.~R.~Park, C.~H.~Kim, J.~Yu {\it et al.},
Phys. Rev. Lett. {\bf 107} 15 156803 (2011)

\bibitem{montory}K.~T.~Delaney, M.~Mostovoy,  N.~Spaldin,
Phys. Rev. Lett. {\bf 102} 15 157204 (2009)

\bibitem{okamura-skyrm}Y.~Okamura {\it et al.}, Nature Comm. 4., 2391 (2013)

\bibitem{heinze}S.~Heinze {\it et al.}, Nature Physics {\bf 7}, 713 (2011)

\bibitem{romming}N.~Romming {\it et al.}, Science {\bf 341}, 636 (2013)


\end{thebibliography}
\end{document}